\pdfoutput=1

\documentclass[12pt]{article}
\usepackage{jheppub}
\usepackage{amsmath,amssymb,array,calc,epsfig,rotating,bm}
\allowdisplaybreaks
\allowdisplaybreaks
\usepackage{graphicx,color,slashed,braket}
\usepackage{bbm}
\usepackage{pifont}
\usepackage{ifpdf}
\usepackage{comment}
\usepackage{float}
\newcommand{\ie}{{\it i.e.,}\ }
\newcommand{\eg}{{\it e.g.,}\ }

\newcommand{\be}{\begin{equation}}
\newcommand{\ee}{\end{equation}}
\newcommand{\ba}{\begin{eqnarray}}
\newcommand{\ea}{\end{eqnarray}}

\def\Im{\mathO$p${\rm Im}\nolimits}

\newcommand{\Sera}[1]{\textbf{\textcolor{cyan}{(#1 - Sera)}}}

\newcommand\nxt[1]  {\\\fnxt#1}

\def\tmoda {\Theta^{\text{(I)}}}

\def\calc         {{\cal C}}
\def\cald         {{\cal D}}

\def\call         {{\cal L}}
\def\calm         {{\cal M}}
\def\caln         {{\cal N}}
\def\calo         {{\cal O}}

\def\calt         {{\cal T}}

\def\calw         {{\cal W}}

\def\complex      {{\mathbb C}}

\def\dd{\delta}
\def\del          {\partial}

\def\ee           {{\rm e}}

\def\Im           {{\rm Im\hskip0.1em}}

\def\sqr#1#2{{\vcenter{\vbox{\hrule height.#2pt
 \hbox{\vrule width.#2pt height#1pt \kern#1pt
 \vrule width.#2pt}\hrule height.#2pt}}}}

\title{Holographic Charged Transport with Higher Derivatives}

\author[a]{Alex Buchel,}
\emailAdd{abuchel@uwo.ca}
\affiliation[a]{Department of Physics and Astronomy, 
University of Western Ontario, London, Ontario N6A 5B7, Canada}
\author[b]{Sera Cremonini,}
\emailAdd{cremonini@lehigh.edu}
\affiliation[b]{Department of Physics, Lehigh University, Bethlehem, PA, 18015, USA}
\author[b]{Mohammad Moezzi,}\emailAdd{mom323@lehigh.edu}
\author[b]{and George Tringas}\emailAdd{georgios.tringas@lehigh.edu}

\notoc

\abstract{We compute the first-order hydrodynamic transport coefficients (shear viscosity
$\eta$, bulk viscosity $\zeta$, and charge conductivity $\sigma$)
for a broad class of  strongly coupled, four-dimensional charged
relativistic gauge theory plasma with holographic
gravitational duals containing higher-derivative corrections.
The landscape of our holographic models captures non-conformal gauge theories
with an arbitrary number of relevant coupling constants and a general scalar potential in the gravitational dual, 
allowing for a systematic exploration of charged transport along generic holographic RG flows.
The leading-order higher-derivative corrections 
probe gauge theories with non-equal central charges $c\ne a$ at the ultraviolet fixed point, and 
enable the engineering of 
 diverse temperature and charge density profiles
for the viscosities and the conductivity. 
Our results establish the 
{\it membrane paradigm} in higher-derivative holographic models:
all the transport coefficients are extracted from the 
black brane horizon values of the gravitational scalars,
and various functions defining the gravitational holographic dual.
}

\begin{document}

\vspace*{-2cm} 
\begin{flushright}
\end{flushright}

\maketitle

\newpage

{\hypersetup{hidelinks}
\tableofcontents
}

\newpage

\section{Introduction}\label{sec:introduction}

Hydrodynamics is a universal effective description of interacting systems close to
thermal and chemical equilibrium at energy scales  much lower than
the local temperature and the chemical potential, as well as any
other microscopic scale of a system. It became an indispensable tool in understanding the
physics of heavy ion collisions at RHIC and LHC, and in particular revealing
a new phase of nuclear matter, known as the strongly coupled quark-gluon plasma (sQGP)
\cite{Shuryak:2004cy,Heinz:2008tv}. Modern relativistic hydrodynamics \cite{Kovtun:2012rj}
is understood
as theory of conservation laws for the stress-energy tensor $T^{\mu\nu}$ and (in the case of
a charged plasma) the $U(1)$ current $J^\mu$,
\begin{equation}
\del_\mu T^{\mu\nu}=0\,,\qquad \del_\mu J^\mu=0 \, ,
\label{conlaw}
\end{equation}
each being a functional of (in general all-orders) gradients of
the fluid local four-velocity $u^{\mu}$, the temperature $T$,
and the chemical potential $\mu$, supplemented by assumptions
about the dissipative response and thermodynamic information such as the equation of
state, relating the energy density $\epsilon$, the pressure $P$
and the charge density $\rho$. 
To leading
order\footnote{Higher-order operators are important for the
stability and causality of (\ref{conlaw}) \cite{Romatschke:2009im}.}
in the gradient expansion, in the Landau
frame\footnote{The first-order transport coefficients discussed in this paper
are frame-independent \cite{Kovtun:2019hdm}.},
it is completely specified by the shear and bulk viscosities ($\eta$ and $\zeta$ correspondingly)
and the charge conductivity $\sigma$: 
\begin{align}\label{order11}
&T^{\mu\nu}=\epsilon u^\mu u^\nu + P\Delta^{\mu\nu}+\Pi^{\mu\nu}_{(1)}+\cdots\,,\qquad \Delta^{\mu\nu}
=\eta^{\mu\nu}+u^\mu u^\nu\,,\qquad \Pi^\mu_{\ \nu} u^\nu=0 \, ,\\
&\Pi^{\mu \nu}_{(1)}=-\eta \sigma^{\mu \nu} -\zeta \Delta^{\mu \nu} (\partial_{\alpha} u^{\alpha})\,, 
\qquad \sigma^{\mu \nu}=\Delta^{\mu \alpha} \Delta^{\nu \beta}
(\partial_{\alpha}u_{\beta}+\partial_{\beta}u_{\alpha})
-\frac 23 \Delta^{\mu \nu}\Delta^{\alpha \beta}(\partial_{\gamma}u^{\gamma})\,, \nonumber
\end{align}
and
\begin{equation}
\begin{split}
&J^{\mu}=\rho u^{\mu} +\nu^{\mu}_{(1)}+\cdots\,,\qquad u^{\mu}\nu_{\mu}=0\,,\qquad
\nu^{\mu}_{(1)}=\sigma\Delta^{\mu \nu}\left(-\partial_{\nu}\mu
+\frac{\mu}{T}\partial_{\nu}T\right) \, ,
\end{split}
\label{order12}
\end{equation}
where $\cdots$ in (\ref{order11}) and (\ref{order12}) indicate dependence on the
second and higher-order gradients, along with higher-order transport coefficients.
While the transport coefficients $\{\eta,\zeta,\sigma\}$ are completely determined
from the microscopic definition of a theory, with a fully prescribed dependence on
the temperature and the chemical potential $\{T,\mu\}$ of a thermodynamic equilibrium state,
such  computations are extremely challenging in practice: hydrodynamics is typically
applied to {\it fluids} which are intrinsically strongly coupled systems. 
The issue of strong coupling is of practical importance to the sQGP:
for finite-temperature thermal gauge theories, computations based on the Boltzmann equation in the
regime of weak coupling $g\ll 1$ predict that the ratio of the shear viscosity to the entropy density
behaves as \cite{Arnold:2000dr,Arnold:2003zc}
\begin{equation}
\frac {\eta}{s}\ \sim\ \frac{\rm const}{g^4 \ln 1/g^2}\ \gg\ 1 \, ,
\label{weakc}
\end{equation}
while the values preferred from RHIC and LHC experiments \cite{Heinz:2011kt}
are close to the universal holographic result at infinitely-strong coupling
\cite{Policastro:2001yc,Buchel:2003tz}
\begin{equation}
\frac{\eta}{s}=\frac{1}{4\pi} \, .
\label{uni}
\end{equation}
For more than two decades
the holographic correspondence \cite{Maldacena:1997re,Aharony:1999ti} has been
the go-to framework to compute transport coefficients of gauge theories at strong coupling ---
the first ever computation of the non-trivial\footnote{The theory here
has to be non-conformal.} bulk viscosity in any gauge theory (weak or strong coupling)
was done in a holographic model \cite{Benincasa:2005iv}.

Although typical top-down holographic models, such as $\caln=4$ superconformal Yang-Mills theory
\cite{Maldacena:1997re}, or its non-conformal cousins, the $\caln=2^*$ gauge
theory \cite{Buchel:2000cn} and the cascading gauge theory \cite{Klebanov:2000hb},
mimic the  general features of the thermodynamics of QCD at the
deconfinement crossover, they fail to reproduce precision finite
temperature lattice data. 
An alternative approach, the {\it Improved Holographic QCD} (ihQCD) 
\cite{Gursoy:2007cb,Gursoy:2007er}, has been
to engineer semi-phenomenological holographic renormalization group
(RG) flows\footnote{In practice, this means making a judicious choice for a
dilaton potential in five-dimensional Einstein-dilaton gravity
with a negative cosmological constant.} to reproduce
the static properties of QCD, such as zero temperature glueball spectra, exact $\beta$-function,
lattice thermodynamics (EoS), etc, and then use the resulting best-fit model to make dynamical
predictions, hopefully of relevance to the sQGP. 
The process of identifying the holographic model that provides the best fit to lattice EoS
can be made algorithmic with the aid of
physics-informed neural networks
\cite{Bea:2024xgv}\footnote{See also \cite{Hashimoto:2021ihd,Hashimoto:2022eij,Chen:2024ckb}.}.
With future improvements in data on the QGP transport coefficients, it is desirable to
use experimental results for $\{\eta,\zeta,\sigma\}$ as an additional input
to the lattice EoS for the engineering of the ihQCD. To achieve such a goal, one needs
an efficient method to compute the transport coefficients in a generic holographic model, ideally
with little overhead on top of the computation of the black brane background
geometry encoding the equation of state of the boundary gauge theory plasma. 
For a restrictive class of two-derivative holographic models, 
an efficient approach
has been known for some time: one can use the membrane paradigm \cite{Buchel:2004qq,Iqbal:2008by}
to compute the shear viscosity $\eta$ and the charge conductivity $\sigma$,
and the Eling-Oz formula \cite{Eling:2011ms,Buchel:2011yv,Buchel:2011wx}
to compute the bulk viscosity $\zeta$. 
Unfortunately, two-derivative holographic
models cannot adequately capture the full phenomenology of the sQGP: for homogeneous and isotropic thermal equilibrium states
of charged plasma they predict \cite{Buchel:2003tz,Benincasa:2006fu}
the universal ratio for the shear
viscosity (\ref{uni}),
which, in particular, is temperature independent,
in contradiction with experimental results \cite{Heinz:2011kt}.  

Until recently, efficient computations of the transport coefficients in holographic models with
higher-derivative corrections were extremely involved
technically: the old membrane paradigm picture of \cite{Iqbal:2008by} needed
modification, and the Eling-Oz formula was found to be simply
incorrect \cite{Buchel:2023fst,Buchel:2024umq}. Rather, the computations
proceeded on a-model-specific-basis,
with the evaluation of the higher-derivative black brane quasinormal modes
\cite{Benincasa:2005qc,Buchel:2018ttd}
or (in the case of the shear viscosity) the two-point boundary correlation function of the
stress-energy tensor in its background
\cite{Buchel:2004di,Cremonini:2009ih,Myers:2009ij,Buchel:2010wf} (see e.g. \cite{Cremonini:2011iq} for a review). 

The issues just mentioned have been resolved with the implementation of the
novel unified framework of \cite{Buchel:2023fst}, based on the extension of
the membrane paradigm method introduced in \cite{Gubser:2008sz}
to holographic models with higher-derivatives. 
The key simplifications behind the analysis of \cite{Buchel:2023fst} resulted from identifying 
a radially conserved quantity even in the presence of higher-derivative terms, which in turn implied that the shear and bulk viscosity can be extracted from the horizon data of the geometry.
Indeed, the highlight of \cite{Buchel:2023fst} was that, for a large class of
higher-derivative holographic RG flows, there was no overhead whatsoever
to evaluate the shear viscosity, once the background black brane geometry
was known. For the bulk viscosity, one needs to solve an additional
master system of reparametrization invariant equations for the
scalar fluctuations, a step which is substantially simpler than solving the corresponding
quasinormal mode equations. 
Furthermore, the results for the bulk viscosity,
much like the ones for the shear viscosity, are obtained
within the developed extended membrane paradigm, \ie one needs
only the values of the bulk scalar fluctuations at the black brane horizon.
As a practical application of the computational framework \cite{Buchel:2023fst},
it was demonstrated in \cite{Buchel:2024umq}
how easy it is to avoid the lower causality bound on $\frac{\eta}{s}$
seen in the conformal RG flows of \cite{Brigante:2008gz}, by using 
non-conformal holographic RG flows, 
and moreover, how to make the ratio
$\frac{\eta}{s}$ temperature dependent at leading order (not just 
due to the infinitesimal corrections from the higher-derivative
terms of the holographic model).  
See  also \cite{Apostolidis:2025gnn} for work on combining the framework of \cite{Buchel:2023fst}  with Bayesian analysis of heavy-ion data, and  
 \cite{Tong:2025rxz} for a particular extension of \cite{Buchel:2023fst} to Gauss-Bonnet-scalar gravity.

In this paper we extend the results of \cite{Buchel:2023fst} in various
ways:
\begin{itemize}
\item We consider higher-derivative holographic RG flows with a bulk
$U(1)$ gauge field --- this allows us to consider charged equilibrium
thermal states of the boundary gauge theory plasma, useful to model, \eg 
QGP states with a baryonic chemical potential $\mu_B$. 
\item We present formulas for all the first-order transport coefficients:
the shear $\eta$ and bulk $\zeta$ viscosities, as well as the $U(1)$
charge conductivity $\sigma$ (the thermal conductivity, which describes the response of the heat flow to the temperature gradients, can also be obtained once we have $\sigma$, see \cite{Son:2006em}).
They will now be functions
of the equilibrium temperature $T$ and the chemical potential $\mu$;
of course they will also depend on whatever microscopic scales
our boundary gauge theory has --- all the relevant couplings of the
boundary gauge theory, and the details of its (holographic) RG flow
(as encoded in an arbitrary choice of the bulk scalar fields potential). 
\item We substantially enlarge the class of holographic models
of \cite{Buchel:2023fst} used to compute $\{\eta,\zeta,\sigma\}$
by promoting 
the coefficients of the
higher-derivative operators of a gravitational model to arbitrary functionals
of the bulk scalars.
\item  The transport coefficients are evaluated within the extended
membrane paradigm picture of \cite{Buchel:2023fst}: there is no computational
overhead to evaluate $\eta$ once the background black brane geometry
is known; for the bulk viscosity $\zeta$ (or the
conductivity $\sigma$), as in  \cite{Buchel:2023fst},
one needs to solve a set of master equations for the bulk scalar fields fluctuations (correspondigly the gauge field fluctuation),
and extract the values of these fluctuations at the black brane horizon.
\end{itemize}

One should keep in mind that our setup assumes that the black brane geometry is homogeneous and isotropic, implying that the dual field theory preserves translational and rotational invariance.
However, the flow in the sQGP is inherently anisotropic, and it is therefore important to understand 
how such anisotropies affect the dynamics near the QCD phase transition, and in particular the temperature dependence of $\frac{\eta}{s}$.
When rotational invariance is broken, $\frac{\eta}{s}$ is no longer universal even in two-derivative theories (see \cite{Landsteiner:2007bd,Basu:2011tt,Erdmenger:2011tj,Rebhan:2011vd} for early studies and \cite{Baggioli:2023yvc,Rougemont:2023gfz,Demircik:2024bxd} for more recent discussions).
Extending our framework to incorporate symmetry breaking would therefore be extremely valuable, not only for the sQGP but also more broadly for the rich landscape of emergent phases in strongly correlated materials.
More efficient methods for computing transport coefficients could help identify generic features of strong interactions in the low-temperature regime and shed light on the interplay between different symmetry-breaking mechanisms,
potentially clarifying the origin of any universal behavior or possible bounds.

Another issue worth mentioning is that in the near extremal, low temperature regime, quantum corrections are expected to modify the semi-classical dynamics of charged black branes (large quantum fluctuations develop in the throat of a near-extremal black hole, see \eg \cite{Mertens:2022irh} for a review). 
It is therefore of interest, working within the holographic framework, to examine how such gravitational quantum corrections can impact the transport properties of the dual system (see \eg \cite{Liu:2024gxr,Gouteraux:2025kta,Nian:2025oei,PandoZayas:2025snm,Cremonini:2025yqe,Kanargias:2025vul,Gouteraux:2025exs}).  
For $\frac{\eta}{s}$ in particular \cite{PandoZayas:2025snm,Cremonini:2025yqe,Kanargias:2025vul,Gouteraux:2025exs}, the focus has been on understanding how it deviates from the universal result (\ref{uni}) as the temperature is lowered.
Several questions arise naturally in this context. 
First, one can ask whether other instabilities will keep one from even accessing the extremal regime \cite{Buchel:2025ves,Buchel:2025jup}. 
More generally, one should examine when the quantum effects 
start competing with those due to the higher-derivative corrections studied here.
Finally, an important point was raised in \cite{Emparan:2025sao}: remaining in the range of validity of hydrodynamics while also maintaining significant 
quantum fluctuations pushes the geometry in the sub-Planckian regime, where the spacetime description breaks down. 
Because of this issue, one can not trust gravitational-quantum corrections to transport properties in the deep quantum regime.
In the semi-classical regime in which such quantum corrections \emph{can} be trusted, it is not clear that they would be strong enough to lead to a significant deviation from the effects of higher-derivative corrections. It would be important to understand the interplay of all these effects further. 
We leave these questions to future work.

The rest of the paper is organized as follows. In the next section
we describe the landscape of bottom-up higher-derivative
holographic models used to compute the first-order hydrodynamic
transport coefficients. We present the explicit expressions for
$\{\eta,\zeta,\sigma\}$ and point to the parts in the
paper needed to implement their evaluation. Sections
\ref{MembraneParadigm}---\ref{TransportCoefficients}
collect technical details and the derivation of the results presented in
section \ref{sec:summary}. 
In particular, in section \ref{MembraneParadigm} we review the extended membrane paradigm of
\cite{Buchel:2023fst}, which made the evaluation of the transport coefficients feasible.
We discuss the black brane gravitational backgrounds (and their
thermodynamics) dual to the equilibrium states of charged holographic QGP
(hQGP) in section \ref{Thermodynamics}. Transport coefficients
are evaluated in section \ref{TransportCoefficients}.
Various technical details, including the agreement test of our charged membrane
paradigm with the results of \cite{Buchel:2010gd}, are delegated to Appendices.

\section{Summary of Results}\label{sec:summary}

We will work with a five-dimensional theory of gravity in AdS coupled to an arbitrary
number of scalars $\phi_i$ and a
single bulk $U(1)$ gauge field, described by 
\begin{align}\label{2der}
S_5&=\frac{1}{16\pi G_N}\int_{\calm_5}d^5x \sqrt{-g}\ \call_5
\\
&\equiv \frac{1}{16\pi G_N}\int_{\calm_5}d^5x \sqrt{-g}\ \biggl[
R+12-\frac 14 \gamma\{\phi_i\}\cdot F^2-\frac{1}{2}\sum_i\left(\del \phi_i\right)^2-V\{\phi_i\}+\dd\call
\biggr]\,, \nonumber
\end{align}
where $\dd\call$ denotes terms involving higher-derivative corrections to Einstein gravity.
In particular, in this paper we consider two classes of models\footnote{Throughout the paper we keep the superscripts $ ^{(I)}$ or $ ^{(II)}$ in
reference to models \eqref{HDLag1} and \eqref{action2}.
The asymptotic AdS radius is set $L=1$, in the absence of the higher-derivative corrections.}:
\begin{itemize}
\item {\bf Model I} ---
perturbative (to leading order in $\beta$)
 four-derivative curvature and gauge field corrections described
by:
\begin{equation} \label{HDLag1}
\begin{split}
\delta\mathcal{L}^{(I)}\equiv & \beta\biggl[\; \alpha_1 \, R^2 +\alpha_2  \, R_{\mu\nu} R^{\mu\nu} + \alpha_3 \, R_{\mu\nu\rho\sigma}R^{\mu\nu\rho\sigma} 
+\alpha_4 \, R F^2  +\alpha_5 \, R_{\mu \nu}F^{\mu \rho}F^{\nu}_{\rho} \\
&+\alpha_6  \,R_{\mu \nu \rho \sigma}F^{\mu \nu}F^{\rho \sigma} 
+\alpha_7 \, R_{\mu \nu \rho \sigma}F^{\mu \rho}F^{\nu \sigma}+ \alpha_8 \, (F^2)^2 + \alpha_9  \, F^4 
\biggr]\,,
\end{split}
\end{equation}
with $F^4 \equiv F_\alpha^\beta F_\beta^\gamma F_\gamma^\delta F_\delta^\alpha$, and the couplings $\alpha_i=\alpha_i\{\phi_j\}$ $(i=1,\dots,9)$
are arbitrary functionals of the bulk scalars $\phi_j$.
The higher-derivative corrections in \eqref{HDLag1} are the most general
set of four-derivative terms one can write down
without introducing the explicit dependence on the space-time
derivatives of the bulk scalars, \ie terms of the type
$(\del\phi_i)^4$ and $R^{\mu\nu} \del_\mu \phi_i\del_\nu \phi_j$.
For the case of a single scalar field, all such terms can be eliminated with a combination of integration by parts and field redefinitions.
For an arbitrary number of scalars, however, this is not longer the case (see the Appendix of
\cite{Buchel:2008vz} for a discussion of both scenarios).
For simplicity, here we restrict our attention to four-derivative terms of the gravity and gauge field sectors only, and leave additional terms involving derivatives of the scalars to future work.

The first-order hydrodynamic transport coefficients  of Model I are:
\nxt the shear viscosity to entropy density ratio:
\begin{align}
4\pi \frac{\eta}{s}\bigg|_{(I)}=&
         1 + \beta\cdot \biggl[\frac{2}{3} \alpha_3 \left(V -12\right)
        - 2 
        \Bigl(
        \sum_i \partial_i V\cdot  \partial_i \alpha_3
        \Bigr)
        + \frac\calc3\biggl(6 \alpha_6 + 3 \alpha_7 
        + \gamma \alpha_3  \nonumber \\&+ 3 \Bigl(
        \sum_i\partial_i \gamma \cdot \partial_i \alpha_3 
        \Bigr)
        \biggr)\biggr] \,;
\label{etasmod1}
\end{align}
\nxt the bulk viscosity to entropy density ratio:
\begin{align}
&9\pi \frac{\zeta}{s}\bigg|_{(I)}=
\biggr(1- \beta\cdot\biggl[
\frac{2}{3}  (V - 12)[5 \alpha_1 + \alpha_2 - \alpha_3] 
    -{\calc}  \biggl( \frac\gamma3 \; [\alpha_1 + 2 \alpha_2 + 7\alpha_3]
    \nonumber \\&
    + [ 2\alpha_4 +  \alpha_5 + 2 \alpha_6 +  \alpha_7]\biggr)\biggr]
\biggr)\sum_i (z_{i,0})^2 \nonumber \\
&+ \beta \cdot \frac{4 \, (\sum_i z_{i,0} \cdot \partial_i V ) -2 \, \mathcal{C} \, (\sum_i z_{i,0} \cdot \partial_i
        \gamma) }{3 \, (\gamma \, \mathcal{C}+2 V -24) }
         \Bigg[\mathcal{C}
    \Big(\sum_j z_{j,0} \cdot \partial_j \gamma \Big)
    \left[\alpha_1+2\alpha_2+7\alpha_3\right] \nonumber \\
    &+
   {2} 
\Big(\sum_j z_{j,0} \cdot \partial_j V\Big)
    \left[5\alpha_1+\alpha_2-\alpha_3\right]
    +2(V-12)\Big(\sum_j z_{j,0} \cdot
    \partial_j[5\alpha_1+\alpha_2-\alpha_3]\Big) \nonumber
\\&- \gamma \, \mathcal{C} \Big(\sum_j z_{j,0} \cdot \partial_j[\alpha_1
    +2\alpha_2+7\alpha_3]\Big)
-3 \, \mathcal{C} \Big(\sum_j z_{j,0} \cdot \partial_j  [2\alpha_4+\alpha_5
      +2\alpha_6+\alpha_7]\Big) \nonumber
   \\& +6 \, \mathcal{C} [2\alpha_4+\alpha_5
      +2\alpha_6+\alpha_7] 
      \Big( \sum_j z_{j,0} \cdot \partial_j\ln\gamma\Big)\Bigg] \,;
\label{zetasmod1}
\end{align}
\nxt the charge conductivity-per-degree-of-freedom
$\hat\sigma\sim\frac{\sigma}{N^2}$, see \eqref{defhs},
to the entropy-per-degree-of-freedom 
$\hat s\sim \frac{s}{N^2}$,
see \eqref{killrh}, ratio :
\begin{equation}
\begin{split}
&\frac{\hat\sigma }{(\hat s)^{1/3}}\bigg|_{(I)}=
\frac{e^2\gamma}{2}a_0^2\biggl(1+\beta\cdot\biggl[\mathcal{C}\gamma(\alpha_2+4\alpha_3)+
\frac{\mathcal{C}}{3}(2\alpha_4+7\alpha_5+22\alpha_6+11\alpha_7) \\
& +\frac{\mathcal{C}}{\gamma}(16\alpha_8+8\alpha_9)-\frac{V-12}{3\gamma}(20\alpha_4+4\alpha_5+4\alpha_6+2\alpha_7)
-\frac 13\kappa\biggr]\bigg)\,;
\end{split}
\label{sigmasmod1}
\end{equation}
where $\del_i\equiv \frac{\del}{\del\phi_i}$, and the quantity $\calc$ depends on the $U(1)$ charge density $\rho$
and the entropy density $s$ of the boundary hQGP
thermal equilibrium state,
\begin{equation}
\calc\equiv\frac{16\pi^2}{\gamma^2}\ \left(\frac{\rho}{s}\right)^2 \, ,
\label{defc}
\end{equation}
and $\kappa$ is defined as
\begin{equation} \label{kappaintro}
    \begin{split}
  \kappa = & \frac{2}{3} \Big(5 \alpha_1 + \alpha_2 - \alpha_3\Big)\Big(V - 12\Big)
    - \frac{\calc}{3}  \Big( 6 \alpha_4 + 3 \alpha_5 + 6 \alpha_6 + 3 \alpha_7 \\
    &+ \gamma (\alpha_1 + 2 \alpha_2 + 7 \alpha_3) \Big)\,.
    \end{split}
\end{equation}
The weak coupling $e$ is introduced to gauge the global (from the boundary gauge theory
perspective) $U(1)$ symmetry, and thus to properly defined the electromagnetic
response of the charged hQGP, see \cite{Kovtun:2008kx}.

All the quantities in \eqref{etasmod1}-\eqref{kappaintro}  are to be evaluated
at the horizon of the dual black brane solution.
Here $z_{i,0}$ are the values of the gauge invariant
scalar fluctuations, at zero frequency, evaluated at the black brane horizon,
see section \ref{sec:bulkviscosity}, and correspondingly, $a_0$ is the horizon
value of the gauge field fluctuation, see section \ref{sec:conductivity}. While the scalar potential
and its derivatives  can be evaluated at the leading $\calo(\beta^0)$
order of the background black brane solution, the horizon values of the scalars
$z_{i,0}$ and of the gauge field fluctuation $a_0$ must be evaluated including $\calo(\beta)$ corrections. 

Notice that there is a tremendous simplification of \eqref{zetasmod1}
when
\begin{equation}
\{\alpha_1,\alpha_2,\alpha_3\}\{\phi_i\}\equiv \lambda_1\{\phi_i\}\cdot
\{1,-4,1\}\,,\qquad \alpha_{4\cdots 9}\{\phi_i\}\equiv 0 \, , 
\label{mod2limit}
\end{equation}
leading to vanishing of the combinations
\begin{equation}
[5\alpha_1+\alpha_2-\alpha_3]\equiv 0\,,\qquad [\alpha_1+2\alpha_2+7\alpha_3] 
\equiv 0\,.
\label{zeros}
\end{equation}
For the choice of couplings \eqref{mod2limit}, the higher-derivative corrections take the form
\begin{equation}
\delta\call^{(I)}\to \delta\call^{(I)}_{GB}\equiv \beta\cdot\lambda_1\{\phi_i\}\cdot \call_{GB} \, ,
\label{mod2gb}
\end{equation}
where the Gauss-Bonnet combination is
\begin{equation}
\call_{GB}\equiv R^2 - 4 R_{\mu \nu}R^{\mu \nu} + R_{\mu \nu \rho \sigma}R^{\mu \nu \rho \sigma} \, ,
\label{defgb}
\end{equation}
and the bulk viscosity to entropy ratio becomes 
\begin{equation}
9\pi \frac{\zeta}{s}\bigg|_{(I),GB}=\sum_i (z_{i,0})^2 \, .
\label{mod2gbzeta}
\end{equation}
The holographic model $\delta\call^{(I)}_{GB}$ is in fact a two-derivative
one, and therefore can be analyzed for finite $\beta$ --- indeed, it is a special
case of Model II below. 
For vanishing charge density, the simplification of the
bulk viscosity in the GB model 
was observed in \cite{Buchel:2024umq}; we see here that this is
the case for the charged hQGP as well (due to the vanishing
of the second combination in \eqref{zeros}).
Alas, there is no simplification in this model for the shear viscosity
\eqref{etasmod1}.

\item {\bf Model II} ---
finite Gauss-Bonnet corrections $\mathcal{L}_{\text{GB}}$ \eqref{defgb}, but only quadratic in the field strength $F$: 
\begin{equation}
\dd\call^{(II)}\equiv
\lambda_1\{\phi_i\} \cdot \mathcal{L}_{\text{GB}} + \epsilon \, \lambda_2\{\phi_i\} \cdot
\mathcal{L}_{\text{GB}} \cdot F^2 \, .
\label{action2}
\end{equation}
Considering Gauss-Bonnet corrections fully nonlinearly --- \ie, the coupling $\lambda_1$ is not taken to be perturbative ---
allows to explore the interplay between the microcausality of the
boundary gauge theory and its hydrodynamic transport. Specifically, following
\cite{Buchel:2009tt}, microcausality of this model restricts
\begin{equation}
-\frac{7}{72}\ \le \lambda_1\{\phi_i\}\bigg|_{\phi_i=0}\ \le \frac{9}{200} \, .
\label{gbc}
\end{equation}
We treat the remaining higher-derivative corrections perturbatively, working to linear
order in $\epsilon$. The motivation behind this particular setup is that
it might provide a useful framework for phenomenological applications to the QGP, in
the regime of small baryonic chemical potential.

The first-order hydrodynamic transport coefficients  of Model II are:
\nxt the shear viscosity to entropy density ratio:
\begin{equation}
\begin{split}
&4\pi \frac{\eta}{s}\bigg|_{(II)}= 1 +  \frac{2}{3}\biggl(\mathcal{D}\lambda_1 -3 \, \sum_i \partial_i\lambda_1 \cdot \mathcal{D}_i \biggr) +\frac83 \, \mathcal{D}^2 \sum_i (\partial_i\lambda_1)^2\\
    &+\frac{2 \, \calc}{3} \epsilon\cdot\biggl[
    \,  \biggl( -\frac{12\lambda_2}{\gamma} \, \sum_i \partial_i \gamma \cdot \mathcal{D}_i
    +10 \mathcal{D} \, \lambda_2 + 6 \sum_i \partial_i \lambda_2 \cdot \mathcal{D}_i\biggr)\\
    &+\frac{16}{3}\,  \mathcal{D}^2 \, \biggl(
    -3 \sum_i\partial_i\lambda_1  \cdot \partial_i\lambda_2-\lambda_1\lambda_2
    +3\lambda_2\sum_i   \partial_i\lambda_1\cdot \partial_i\ln \gamma   \biggl)  \\&
    -\frac{128}{3} \,  \mathcal{D}^3 \, \lambda_2 \,  \sum_i (\partial_i\lambda_1)^2 \biggr] \, ,
\end{split}
\label{etasmod2}
\end{equation}
where
\begin{equation}
\cald\equiv V-12+\frac \gamma2\, \calc\,,\qquad \cald_i\equiv \del_i V
-\frac{\calc}{2}\,\del_i\gamma \, ;
\label{defds}
\end{equation}
\nxt the bulk viscosity to entropy density ratio\footnote{While there is no explicit dependence on  $\lambda_i$
in \eqref{zetasmod2} --- much like in \eqref{mod2gbzeta},
see also \cite{Buchel:2024umq}  --- both
couplings enter in the equations for the scalar fluctuations
$z_{0,i}$. }: 
\begin{equation}
\begin{split}
&9\pi \frac{\zeta}{s}\bigg|_{(II)}=\sum_i z_{i,0}^2\,;
\end{split}
\label{zetasmod2}
\end{equation}
\nxt the charge conductivity-per-degree-of-freedom to the entropy-per-degree-of-freedom
 ratio: 
\begin{equation}
\begin{split}
\frac{\hat\sigma}{(\hat s)^{1/3}}\bigg|_{(II)}=&
 \frac{e^2 \gamma}{2 } \, a_0^2 \, \biggl(1
    -\epsilon\cdot \frac{4 \, (2 V-24 + \gamma \mathcal{C})^2}{3\gamma}\lambda_2 \bigg) \, .
\end{split}
\label{sigmasmod2}
\end{equation}
Once again, $\calc$ is given by \eqref{defc},  all the quantifies in
\eqref{etasmod2}-\eqref{sigmasmod2}
are to be evaluated at the horizon of the
dual black brane solution, including the values of the
 zero-frequency gauge invariant scalars $z_{i,0}$
(see section \ref{sec:bulkviscosity}) and the gauge field $a_0$ (see section \ref{sec:conductivity}) fluctuations.

\end{itemize}

\section{Membrane Paradigm in Higher-Derivative Holography}\label{MembraneParadigm}

In holography, transport coefficients such as the shear and bulk viscosities $\eta$ and $\zeta$,  
can be extracted in several complementary ways. These include computing stress-energy tensor correlators via Kubo formulas, or analyzing linearized quasinormal modes of black brane backgrounds, which in the hydrodynamic limit correspond to the shear and sound modes of the dual field theory. According to the standard holographic prescription, such correlators are obtained from the near-boundary behavior of fluctuating bulk fields, subject to appropriate boundary conditions at the horizon.

Hydrodynamics, however, is an effective description valid at long wavelengths and low frequencies, and one therefore expects it to be encoded in the infrared properties of the bulk geometry -- namely, in the near-horizon region and its fluctuations. This naturally raises the question of to what extent the horizon of a black brane fully captures the hydrodynamic behavior and transport properties of a strongly coupled plasma. Indeed, diffusive modes can be understood 
\cite{Kovtun:2003wp,Starinets:2008fb} as fluctuations of the black brane horizon within the membrane paradigm 
\cite{Damour:1979wya,Thorne:1986iy},
which models the horizon as a fictitious fluid. This perspective was sharpened in  \cite{Buchel:2004qq,Iqbal:2008by}, leading to several prescriptions for computing transport coefficients directly from the  horizon data. Nevertheless, such approaches have so far been restricted to special classes of theories.

A central point we emphasize in this paper is that the quantities required to compute transport coefficients such as $\eta$, $\zeta$ and $\sigma$ can, in fact, be obtained generically from radially conserved currents, even in the presence of higher-derivative corrections. As a consequence, these coefficients can be evaluated entirely at the black brane horizon.
Our goal in this section is to review the argument that explains the origin of the conserved current, as we do next.

Let's denote schematically the set of fluctuations
of the metric, gauge field and scalar fields in the theory, respectively, by
${h(t,r),a(t,r),\varphi(t,r)}$. The set of fluctuations will need to be adapted to the particular matter content of the model of interest, but the logic is general and thus remains the same.
Starting from the quadratic effective action for the fluctuations, and 
assuming that the harmonic form of each perturbation  is given by  
$h(t,r)=e^{- i w t} \, h_{w}(r)$ (with analogous expressions for all the other fluctuations),  we can easily construct a \emph{complexified} version of the effective action,
as originally
done in \cite{Gubser:2008sz}.
Such complexified action is then   
a functional of $h_{w}(r)$ and $h_{w}^*(r)\equiv h_{-w}(r)$, and similarly for the remaining perturbation, 
\begin{equation}
S^{(2)}= \frac{1}{16\pi G_N}\int_0^{r_h} dr\ \call_{\complex}\{h_{w},h^*_{w},a_{w},a^*_{w},\varphi_{w},\varphi^*_{w}\}\,.
\label{h12action}
\end{equation}
The action is constructed  so that the 
equations of motion for the fluctuations, given by 
\begin{equation}
\frac{\delta S^{(2)}}{\delta h^*_{w}} = 0 \,, \quad 
\frac{\delta S^{(2)}}{\delta a^*_{w}} =0\, , \quad \frac{\delta S^{(2)}}{\delta \varphi^*_{w}} =0 \, ,
\label{h12eom}
\end{equation}
are identical to the ones obtained from the original, uncomplexified effective action. 

One can then show that, 
on shell, the Lagrangian $\call_{\complex}$ appearing in 
the (complexified) effective action for the fluctuations  can be re-expressed as a total derivative, \ie  schematically as  
\begin{equation}
\call_{\complex}=16\pi G_N\left( h_{w}^*\cdot \frac{\dd S^{(2)}}{\dd h^*_{w}} + a_w^* \cdot \frac{\dd S^{(2)}}{\dd a^*_{w}}  + \varphi_{w}^*\cdot \frac{\dd S^{(2)}}{\dd \varphi^*_{w}} + \ldots \right)+\del_r J_w\,.
\label{totlc2}
\end{equation}
Here $J_w$  denotes the current, whose explicit form depends on the model under consideration. 
For a theory with higher-derivatives, as in the models we study here, the current will involve contributions coming from the leading two-derivative terms in the action, as well as corrections coming from the higher-derivative terms.

The key observation  originally made in \cite{Gubser:2008sz}
was that, on shell, the imaginary part of $J_w$ in \emph{two-derivative} holographic models
 is radially conserved.
 This was verified  explicitly in \cite{Buchel:2023fst} in a simpler set of theories with higher-derivative corrections, as well as in our more general setup in this paper. 
However, the crucial point is that this property holds in general,   
\begin{equation}
\del_r (J_w-J_{-w})\bigg|_{on-shell}=2\ \del_r \Im J_w\bigg|_{on-shell} =0\,, \label{conscurrent}
\end{equation}
independently of the matter content and higher-derivative terms.

Indeed, the conservation law \eqref{conscurrent} is a direct consequence
of the exact $U(1)$ symmetry of the complexified Lagrangian $\call_\complex$  
that
rotates the phase of  fluctuations, namely $h_{w}\to e^{i\theta} h_{w}$ and
$h_{-w}\to e^{-i\theta} h_{-w}$, for all fluctuations under consideration.
Since each fluctuation is complexified in the same way, \ie all fluctuations share the same phase $\theta$, the argument is entirely general.
The conserved Noether charge associated
with this symmetry is precisely $\Im J_w$.
Indeed, 
\begin{equation}
16\pi G_N\ \frac{\delta S^{(2)}}{\delta \theta} = -i\ \del_r\theta\cdot
(J_w-J_{-w})\,,
\label{conjw2}
\end{equation}
for infinitesimal $\theta$-rotations. In \cite{Gubser:2008sz}
this conserved charge was interpreted as the radially conserved number flux of gravitons\footnote{See \cite{Gubser:2008sz} for further discussion and related earlier work.}$^,$\footnote{It is an interesting open question as to why the quadratic action for the fluctuations has this peculiar property, \eqref{totlc2}.}.

As shown in  \cite{Buchel:2023fst} (see also
section \ref{TransportCoefficients} below), the Kubo formulas for the
transport coefficients require evaluations of the corresponding currents
$\Im J_w$ at the asymptotically $AdS_5$ boundary. Since $\Im J_w$
is a holographic RG flow invariant \eqref{conscurrent}, it might as
well be evaluated at the regular 
black brane
horizon (dual to a thermal equilibrium state of hQGP),
implementing the {\it membrane paradigm} 
in this more general class of theories, 
and
ultimately leading to the results reported in section \ref{sec:summary}.

\section{Black Branes and the Boundary hQGP
Thermodynamics}\label{Thermodynamics}

We are ready to  set the stage for our computation of thermodynamic and hydrodynamic properties of the hQGP
dual to the gravitational Model I, see \eqref{2der} and \eqref{HDLag1},  and  the gravitational Model II, see
\eqref{2der} and  \eqref{action2}. Our setup builds directly on that of \cite{Buchel:2023fst},  expanding it in two crucial ways: one is by considering the presence of a $U(1)$ gauge field, which allows us to probe systems at finite charge density,  and the other by working with couplings that are scalar dependent, leading to a much broader framework for phenomenological applications to \eg QCD.

\subsection{Model I}\label{sec:modelI}

The first holographic model we consider involves gravity in five dimensional asymptotically AdS space,
coupled to  a $U(1)$ gauge field $A$ and an arbitrary number of scalar fields $\phi_i$. 
We will work with the effective action given by \eqref{2der} and \eqref{HDLag1},
\begin{equation}
\begin{split}
&S_{(I)}
\equiv \frac{1}{16\pi G_N}\int_{\calm_5}d^5x \sqrt{-g}\ \biggl[
R+12-\frac 14 \gamma\{\phi_i\}\cdot F^2-\frac{1}{2}\sum_i\left(\del \phi_i\right)^2-V\{\phi_i\}+\dd\call^{(I)}
\biggr]\,,
\end{split}
\label{action1}
\end{equation}
and consider charged black brane solutions in this theory -- dual to a thermal equilibrium state at finite chemical potential. We take the metric, the bulk gauge field,  and the matter fields to be given by 
\begin{equation}\label{metric}
\begin{split}
    ds_5^2 &= - c_1^2(r)\, dt^2 + c_2^2(r)  \, d\boldsymbol{x}^2 + c_3^2(r) \, dr^2 \, , \\ 
    A_\mu &= \Bigl(A(r),0,0,0,0\Bigr)\, , \quad \quad\,\,\,\, \phi_i = \phi_i (r) \,,
\end{split}
\end{equation}
where  
%
%
$r \in \left[ 0,r_h \right]$ is the holographic radial coordinate, with $r_h$ the regular black brane horizon radius, identified by
\begin{equation}
    \lim_{r \rightarrow r_h} c_1 =0\,,
\end{equation}
and $r=0$ is the asymptotic $AdS_5$ boundary.
After evaluating the effective action \eqref{action1} on the background ansatz (\ref{metric}),
one finds an effective one-dimensional action of the form
\begin{equation} \label{effActionS1}
    S_{(I)} = \frac{1}{16 \pi G_N} \int_{0}^{r_h} dr \left[ \mathcal{I}^{\text{(I)}} + \beta \cdot \delta\mathcal{I}^{\text{(I)}}\right],
\end{equation}
from which one can easily obtain the equations of motions (EoMs) for the background fields --- $\mathcal{I}^{\text{(I)}}$ and $\delta\mathcal{I}^{\text{(I)}}$ are given in \eqref{effBGAction1}-\eqref{effBGAction2}.
In particular, the equations of motion for the metric and scalar fields are given by (we introduce $\del_i\equiv
\frac{\del}{\del \phi_i}$)  
\begin{align} \label{BGEOMS3}
     &c_1'' + \frac{1}{6} c_1 c_3^2 (V - 12) + \frac{c_1}{12} \, \sum_{i}(\phi_i')^2 + \frac{2 c_1' c_2'}{c_2} - \frac{c_1 c_2'^2}{c_2^2} - \frac{c_1' c_3'}{c_3} - \frac{5 \gamma A'^2}{12 c_1} + \beta  \cdot  \Big[ \ldots  \Big] =0 \,, \nonumber \\
     &c_2'' + \frac{1}{6} c_2 c_3^2 (V - 12) + \frac{c_2}{12} \, \sum_{i}(\phi_i')^2 + \frac{c_2'^2}{c_2} - \frac{c_2' c_3'}{c_3} + \frac{c_2 \gamma A'^2}{12 c_1^2}  + \beta  \cdot  \Big[ \ldots  \Big] =0 \,, \nonumber \\
     &\sum_{i}(\phi_i')^2 - 2 c_3^2 (V -12) - \frac{12 c_1' c_2'}{c_1 c_2} - \frac{12 c_2'^2}{c_2^2} - \frac{\gamma A'^2}{c_1^2}   + \beta  \cdot  \Big[ \ldots  \Big] =0 \,,\nonumber \\
     &\phi_i'' - c_3^2 (\partial_i V) + \frac{c_1' \phi_i'}{c_1} + \frac{3 c_2' \phi_i'}{c_2} - \frac{c_3' \phi_i'}{c_3} +\frac{(\partial_i \gamma) A'^2}{2 c_1^2}   + \beta  \cdot  \Big[ \ldots  \Big] =0 \, ,
\end{align}
where primes $'$ denote derivatives with respect to $r$, and the $\mathcal{O}(\beta)$ contributions have been omitted for readability, but are  included in the analysis throughout.
Moreover, the EoM for the gauge field $A$ can be written as a total derivative, 
\begin{equation} \label{BGEOMS1}
    \frac{d}{dr} \bigg( - \frac{c_2^3 \gamma A'}{c_1 c_3} + \beta \, \delta_\mathcal{A}^{\text{(I)}}\bigg) =0 \, , 
\end{equation}
where $\delta_\mathcal{A}^{\text{(I)}}$  denotes the contribution from the higher-derivative terms,
\begin{align} \label{BGEOMS2}
        \delta_\mathcal{A}^{\text{(I)}} &= \;\alpha_4 \bigg( -\frac{24 c_2^2 c_1' c_2' A'}{c_1^2 c_3^3} - \frac{24 c_2 c_2'^2 A'}{c_1 c_3^3} + \frac{8 c_2^3 c_1' c_3' A'}{c_1^2 c_3^4} + \frac{24 c_2^2 c_2' c_3' A'}{c_1 c_3^4} - \frac{8 c_2^3 c_1'' A'}{c_1^2 c_3^3} \nonumber \\
        &- \frac{24 c_2^2 c_2''A'}{c_1 c_3^3}\bigg) 
        +  \alpha_5\bigg( - \frac{6 c_2^2 c_1' c_2' A'}{c_1^2 c_3^3} + \frac{4 c_2^3 c_1' c_3' A'}{c_1^2 c_3^4} + \frac{6 c_2^2 c_2' c_3' A'}{c_1 c_3^4} - \frac{4 c_2^3 c_1'' A'}{c_1^2 c_3^3} - \frac{6 c_2^2 c_2'' A'}{c_1 c_3^3} \bigg) \nonumber \\
        &+\big( 2 \alpha_6 +\alpha_7 \big) \bigg( \frac{4 c_2^3 c_1' c_3 ' A'}{c_1^2 c_3^4} - \frac{4 c_2^3 c_1'' A'}{c_1^2 c_3^3} \bigg) +\big( 2 \alpha_8 +\alpha_9 \big)\bigg( - \frac{8 c_2^3 A'^3}{c_1^3 c_3^3}\bigg) \, .
\end{align}
By integrating \eqref{BGEOMS1}, we then 
immediately conclude that the  following expression must be the RG flow constant, 
\begin{equation}
\label{BGEOMS1b}
    - \frac{c_2^3 \gamma A'}{c_1 c_3} +  \beta \, \delta_\mathcal{A}^{\text{(I)}} = \text{constant} \equiv \Theta^{\text{(I)}} \, .
\end{equation}

Now that we have introduced our background geometry, we move on to computing the 
 temperature $T$ and entropy density $s$ of the
boundary thermal state. 
In holography, the temperature $T$ can be calculated from the analytical continuation of the black brane geometry (\ref{metric}) by requiring the conical deficit angle to vanish. We have
\begin{equation} \label{Temp}
    2 \pi T = \lim_{r \rightarrow r_h} \left[ - \frac{c_2}{c_3} \left( \frac{c_1}{c_2}\right)' \right] =  \lim_{r \rightarrow r_h} \left[ - \frac{c_1'}{c_3} + \frac{c_1 c_2'}{c_2 c_3} \right] = \lim_{r \rightarrow r_h} \left[ - \frac{c_1'}{c_3}\right],
\end{equation}
where we used the fact that $c_1$ vanishes at the horizon. 
Moreover, the thermal entropy density of the boundary field theory is determined by the entropy of the dual black brane \cite{Witten:1998zw}. In Einstein gravity coupled to matter without higher-derivative corrections, this is simply the Bekenstein entropy $s_B$,
\begin{equation} \label{Bekenstein}
    s_B = \lim_{r \rightarrow r_h} \frac{c_2^3}{4 G_{N}}\, .
\end{equation}
In our model, however, because of the presence of  higher-derivative terms the latter needs to be replaced by the Wald entropy $s_{W}$ \cite{Wald:1993nt}, 
\begin{equation}
    s_{W} = - \frac{1}{8 \pi G_{N}} \lim_{r\rightarrow r_h} \left[c_2^3 \, \epsilon_{\mu \nu}\epsilon_{\rho \sigma} \frac{\delta L_5}{\delta R_{\mu \nu \rho \sigma}} \right].
\end{equation}
Alternatively, one can extract the entropy from the Gibbs free energy density of the boundary gauge theory.
This is the approach we adopt here.

According to the holographic principle \cite{Maldacena:1997re}\cite{Aharony:1999ti}, the on-shell gravitational action $S_{(I)}$ (\ref{effActionS1}), properly
renormalized \cite{Skenderis:2002wp}, is identified with the Gibbs free energy density $\Omega$ of the boundary field theory. 
When going on-shell, the integrand of the effective action $S_{(I)}$ can be written as a total derivative. This is a crucial step, which simplifies 
the calculation significantly. Indeed, we find
\begin{equation} \label{BGTotalDer}
    \mathcal{I}^{\text{(I)}} + \beta \cdot \delta\mathcal{I}^{\text{(I)}} =
    -\frac{6c_2^2c_1}{c_3} \cdot \left( \text{EoM for }c_2\right) + \frac{d}{d r} \biggl\{ - \frac{2 c_2^3 c_1'}{c_3} + \beta \cdot \delta_\mathcal{B}^{\text{(I)}}\biggl\} \, - \, A' \, \Theta^{\text{(I)}},
\end{equation}
where $\Theta^{\text{(I)}}$ is defined in (\ref{BGEOMS1b}), and $\delta_\mathcal{B}^{\text{(I)}}$ is given in \eqref{DeltaBMod1}. Using the expression above, $\Omega$
takes the form
\begin{align} \label{FreeEnergy1}
    - \Omega = S_{(I)} \Bigg\rvert_{\text{on-shell}} = \, &
    \frac{1}{16\pi G_N} \Biggl[\int_{0}^{r_h} dr \, \frac{d}{dr}\biggl\{ -\frac{2 c_2^3 c_1'}{c_3}  + \beta \cdot \delta_\mathcal{B}^{\text{(I)}}\biggl\} \, - \, \int_0^{r_h} dr \, A' \, \Theta^{\text{(I)}}\Biggl] 
    \nonumber \\ &+\lim_{r\to 0}\ \biggl[S_{GH}+S_{ct}\biggr]\,,
\end{align}
where $S_{GH}$ is a generalized Gibbons-Hawking term \cite{Buchel:2004di,Cremonini:2009ih} 
necessary to have a well-defined variational principle, and $S_{ct}$ is the counter-term
action. However, we will not need the explicit form of either of these corrections.
Because $\tmoda$ is a holographic RG flow invariant  \eqref{BGEOMS1},
we can explicitly evaluate the second integral in \eqref{FreeEnergy1}:
\begin{equation}
\int_0^{r_h} dr \, A' \, \Theta^{\text{(I)}}=\tmoda \int_0^{r_h} dr \, A'=\tmoda\cdot
\left (A\bigg|^{r\to r_h}-A\bigg|_{r\to 0}\right)=-\mu\, \tmoda \, ,
\label{2ndint}
\end{equation}
where we used the fact that the AdS boundary condition of the gauge potential $A$
determines the chemical potential $\mu$ of the boundary  theory, while $A$ vanishes
at the horizon of black brane,
\begin{equation}
\lim_{r\to 0 } A=\mu\,,\qquad \lim_{r\to r_h} A=0 \, .
\label{defmu}
\end{equation}
We can rearrange \eqref{FreeEnergy1} to explicitly implement
the basic thermodynamic relation (here $\rho$ is the $U(1)$
charge density),
\begin{equation}
- \Omega =s\, T+\mu\, \rho - \mathcal{E} \, ,
\label{basthemo}
\end{equation}
leading to
\begin{equation} \label{FreeEnergy2}
\begin{split}
     - \Omega = \frac{1}{16\pi G_N} \,\Biggl[\lim_{r \rightarrow r_h} \bigg[ - \frac{2 c_2^3 c_1'}{c_3} + &\beta\cdot\delta_\mathcal{B}^{\text{(I)}}\bigg] +\mu\, \tmoda\\
     &- \lim_{r \rightarrow 0} \bigg[  - \frac{2 c_2^3 c_1'}{c_3} + \beta \cdot \delta_\mathcal{B}^{\text{(I)}} + S_{\text{GH}} + S_{\text{ct}}\bigg]\Biggl] \, .
\end{split} 
\end{equation}
Comparing \eqref{basthemo} and \eqref{FreeEnergy2} we
identify\footnote{Of course, this is correct up to an arbitrary constant;
but comparing with the thermal AdS, we find that this constant must be zero.
We verified the identification of $\tmoda$ with the charge density
$\rho$ by carefully matching our formalism to STU models thermodynamics
\cite{Behrndt:1998jd}.
}
\begin{equation}\label{defstrho}
    s T \equiv s_W \, T = \, \frac{1}{16\pi G_N}\,\lim_{r \rightarrow r_h} \bigg[ - \frac{2 c_2^3 c_1'}{c_3} + \beta \cdot \delta_\mathcal{B}^{\text{(I)}}\bigg]\,,
    \qquad \rho=\frac{1}{16\pi G_N}\,\tmoda\,.
\end{equation}
As expected, in the absence of higher-derivative corrections  this expression reduces to the Bekenstein entropy (\ref{Bekenstein}). 
Later on in our analysis, we will need to use the following ratio of entropies,
\begin{equation}
\label{swsb}
    \frac{s_W}{s_B} \equiv 1 + \beta \cdot \lim_{r \rightarrow r_h} \kappa(r) \,, \quad\quad
    \kappa \equiv - \frac{c_3}{2 c_2^3 c_1'} \cdot \delta_\mathcal{B}^{\text{(I)}}\,.
\end{equation}
The expression for $\delta_\mathcal{B}$ includes various derivatives of the background fields, for which we need to implement the background EoMs (\ref{BGEOMS1})-(\ref{BGEOMS3}) working perturbatively to linear order in $\mathcal{\beta}$, as explained in \cite{Buchel:2023fst}. With the final expression for $\kappa$ in hand, we can explicitly evaluate it at the horizon, as we explain next.

 In order to evaluate various quantities at the horizon, it is convenient to fix the residual diffeomorphism in (\ref{metric}), and redefine
 the gauge potential $A$, as\footnote{The final result is, of course, independent of this choice.}  
\begin{equation} \label{BGfunctions1}
    \begin{split}
     c_1&= \frac{1}{r} f^{1/2} g^{1/2}\,,
     \qquad
     c_2 = \frac{1}{r}\,,\qquad
     c_3 = \frac{1}{r f^{1/2} }\,,\qquad A \equiv  Q \; a \, g^{1/2}\,,
    \end{split}
\end{equation}
where  $Q$ is just a constant, and $a=a(r)$.
To zeroth order in $\beta$, we have the following  equations, 
\begin{equation} \label{BGfunctions2}
    \begin{split}
         & f' = -\frac{4}{r} + \frac{4 f}{r} +\frac{V}{3 r} + \frac{1}{6} r f \sum_i\phi_i'^2 + \frac{1}{6} Q^2 r^3 \gamma a'^2 + \frac{Q^2 r^3 \gamma  g' a a'}{6 g}+ \frac{Q^2 r^3 \gamma g'^2 a^2}{24 a^2} \, , \\
    & a' = -\frac{r}{3} g  \sum_i\phi_i'^2 \, .
    \end{split}
\end{equation}
Furthermore, using (\ref{BGfunctions1})-(\ref{BGfunctions2}), regularity of $\phi_i(r)$ at the horizon implies\footnote{Here and throughout the paper, we use LHS $\sim$ RHS to denote $\lim_{r\rightarrow r_h} (LHS/RHS) =1$.} (again to zeroth order in $\beta$)
\begin{equation} \label{phiHRZN}
     \quad \phi_i' \sim \frac{6 (\partial_i V) - 3  Q^2  r^4(\partial_i \gamma) a'^2}{r \left[ 2 (V - 12) + Q^2\,  r^4 \,  \gamma \, a'^2\right]} \quad\quad \text{as} \quad\quad r \rightarrow r_h \,.
\end{equation}
Using (\ref{BGfunctions2})-(\ref{phiHRZN}), we evaluate $\kappa$ at the horizon and find the ratio of the entropies \eqref{swsb} we are after, 
\begin{equation} \label{sWsBRatio}
    \begin{split}
  \kappa \sim  \frac{2}{3} \Big(5 \alpha_1 + \alpha_2 - \alpha_3\Big)\Big(V - 12\Big)
    - \frac{\calc}{3}  \Big( 6 \alpha_4 + 3 \alpha_5 + 6 \alpha_6 + 3 \alpha_7 + \gamma (\alpha_1 + 2 \alpha_2 + 7 \alpha_3) \Big)\,,
    \end{split}
\end{equation}
where we are introducing
\begin{equation}
\calc\ \sim\ Q^2 r^4 a'^2 \, .
\label{defco}
\end{equation}
Remarkably, $\calc$ can be related to the thermodynamic properties of the
boundary gauge theory equilibrium state (as already noted in \eqref{defc}). Indeed, to
zeroth order in $\beta$, we see from \eqref{BGEOMS1b} that 
\begin{equation}
\label{fun}
\lim_{r\to r_h} \tmoda =\lim_{r\to r_h} -\frac{\gamma Q a'}{r}\, = 16\pi G_N\, \rho \, ,
\end{equation}
where in the first equality we used \eqref{BGfunctions1} and the horizon
boundary condition for $A$ (and hence for $a$) \eqref{defmu};
the second equality comes from the RG invariance of $\tmoda$, and the AdS
boundary identification \eqref{defstrho}.
From \eqref{defco} and \eqref{fun} we conclude
\begin{equation}
\label{fun1}
\calc\,\sim\, (16\pi G_N\, \rho)^2\ (r^3)^2\,  \frac{1}{\gamma^2}\,\sim\,
\frac{16\pi^2}{\gamma^2}\ \left(\frac{\rho}{s}\right)^2 \, ,
\end{equation}
where in the last equality we used, to order $\calo(\beta^0)$,  that $4 G_N\, s\,\sim c_2^3 = r^{-3}$.

Notice that when $\delta \call^{(I)}\equiv \delta \call^{(I)}_{GB}$
(see \eqref{mod2limit} for the choice of couplings), there is no difference between the Wald and
the Bekenstein entropies of the black brane to leading order in
$\beta$ \eqref{sWsBRatio}, \ie
\begin{equation}
\kappa\bigg|_{\rm Model-I,\ GB}\, =\, 0 \, .
\label{k1gb}
\end{equation}
Finally, we can see that if we turn off the background gauge field, \ie set $Q =0$,
we recover from \eqref{sWsBRatio} the corresponding result in \cite{Buchel:2023fst}.

\subsection{Model II}\label{sec:modelII}

In the previous model, we considered a general set of \emph{perturbative} four-derivative corrections \eqref{HDLag1}, working to linear order in the parameter $\beta$.  Interestingly, it is known that for a particular combination of certain curvature corrections --- the Gauss-Bonnet term \eqref{defgb} --- one can obtain a system of equations that can be solved exactly, i.e. even for a \emph{finite} Gauss-Bonnet parameter.  
Precisely because of this reason, 
the well-known Gauss-Bonnet correction 
has been of particular interest as a toy model for non-perturbative physics.

The effective action for Model II is given by \eqref{2der} and \eqref{action2},
\begin{equation}
\label{defmod2f}
      S_{(II)}=\frac{1}{16\pi G_N}\int_{\calm_5} d^5x\sqrt{-g}\biggl[R+12-\frac{1}{4}\gamma\{\phi_i\}
     \cdot  F^2-\frac{1}{2}\sum_{i}(\partial\phi_i)^2-V\{\phi_i\}+\delta\call^{(II)}\biggr] \, .
\end{equation}
Following the same steps as in the previous section, we start with evaluating (\ref{defmod2f})
on the background (\ref{metric}), and obtain the following one-dimensional effective action,
\begin{equation} \label{effActionS2}
    S_{(II)} = \frac{1}{16 \pi G_N} \int_{0}^{r_h} dr \left[ \mathcal{I}^{\text{(II)}} + \epsilon \cdot \delta_{\mathcal{I}}^{\text{(II)}}\right],
\end{equation}
from which we can extract the EoMs for the background fields.
Similarly to Model I, we find that the EoM for the gauge field $A$ can be written as
a total derivative,
\begin{equation} \label{GBBGEOMS1}
    \frac{d}{dr} \bigg( - \frac{c_2^3 \gamma A'}{c_1 c_3} + \epsilon \, \delta_\mathcal{A}^{\text{(II)}}\bigg) =0 \, ,
\end{equation}
which can be integrated to yield
\begin{equation} \label{GBBGEOMS1b}
     - \frac{c_2^3 \gamma A'}{c_1 c_3} +  \epsilon \, \delta_\mathcal{A}^{\text{(II)}} = \text{constant} \equiv \Theta^{\text{(II)}},
\end{equation}
with $\delta_\mathcal{A}^{\text{(II)}}$ given by
\begin{equation} \label{GBBGEOMS2}
    \begin{split}
        \delta_\mathcal{A}^{\text{(II)}} = \frac{96 A' c_2' \lambda_2}{c_1^2 c_3^5} \big( c_2 c_2' c_1'' + c_1 c_2' c_2'' + c_1' c_2'^2 + 2 c_2 c_1' c_2''\big) - \frac{96 A' c_2'^2  c_3' \lambda_2}{c_1^2 c_3^6} \big( 3 c_2 c_1' + c_1 c_2'\big).
    \end{split}
\end{equation}
Corresponding EoMs for other background fields are given in \eqref{BGEOMModelTwo}.
\\

Literally repeating the analysis of section \ref{sec:modelI}, we obtain expressions
for $sT$ and $\rho$ (as in \eqref{defstrho} with appropriate modifications $ ^{(I)}\to ^{(II)}$),
and evaluate the ratio of the Wald entropy $s_W$ to the Bekenstein entropy $s_B$
of the black brane to be 
\begin{equation} \label{sWsBModelTwo}
    \frac{s_W}{s_B}  \sim 1+\epsilon\cdot 0\,,
\end{equation}
\ie to the  leading order in $\epsilon$, there is no difference between the entropies:
$s\equiv s_W=s_B$.

In what follows, we will also need the  expression for $\calc$, introduced  as \eqref{defco}, but
equivalent to \eqref{fun} --- in this case fully nonlinear in $\lambda_1$, but to zeroth
order in $\epsilon$.

\section{Transport Coefficients}\label{TransportCoefficients}

We are now ready to obtain the transport coefficients we are after, for the two models we presented in section \ref{Thermodynamics}. 
We will compute the shear and bulk viscosities $\eta$ and $\zeta$, as well as the conductivity $\sigma$, using Kubo formulas.
In each subsection below, before turning to the detailed computations relevant to the two models, 
we outline the main steps needed to obtain each transport coefficient.

\subsection{Shear Viscosity}\label{ShearViscosity}

Following \cite{Policastro:2001yc,Buchel:2004di}, we use the Kubo formula to compute the shear viscosity from the retarded two-point correlation function of the boundary stress-energy tensor $T_{12}$,
\begin{equation}
\eta=-\lim_{w\to 0}\frac 1w \Im G_R(w)\,,\qquad G_R(w)=-i\int dtd\bm{x} \, e^{i w t}\theta(t)\langle
[T_{12}(t,{\bm{x}}),T_{12}(0,\bm{0})]\rangle\,,
\label{kuboeta}
\end{equation}
where the indices $1,2$ denote two of the spatial directions $x_1,x_2$. 
To compute $G_R(w)$
using holography, we add to the background metric the following perturbation,
\begin{equation}\label{h12}
ds_5^2\ \to\ ds_5^2+2h_{12}(t,r)dx_1 dx_2\,.
\end{equation}
Using symmetry arguments \cite{Policastro:2002se} one can show that all the remaining fluctuations of the
metric, gauge field, and bulk scalars can be consistently set to zero. Moreover,
we can restrict to $SO(3)$ invariant metric perturbations. 

Next, for each model we expand the action to quadratic order in the fluctuation \eqref{h12}, obtaining the quadratic effective action
$S^{(2)}\big\{h_{12}(r,t)\big\}$, as a functional of the metric perturbation
\eqref{h12}.
Assuming that the harmonic dependence of the fluctuation is given by 
\begin{equation}\label{replacemnt1}
    h_{12}(t,r)=e^{-iwt}h_{12,w}(r) \,,\quad\quad
    h_{12}^{*}(t,r)=e^{iwt}h_{12,-w}(r) \,,
\end{equation}
the effective action can then be written in terms of $h_{12,w}(r)$ and $h_{12,w}^*(r)\equiv h_{12,-w}(r)$, 
\begin{equation}
S^{(2)}= \frac{1}{16\pi G_N}\int_0^{r_h} dr\ \call_{\complex}\{h_{12,w},h^*_{12,w}\}\,.
\label{h12actionShearI}
\end{equation}
We should note that the latter is constructed  so that the 
equations of motion for the fluctuation $h_{12,w}$, given by 
\begin{equation}
\frac{\delta S^{(2)}}{\delta h^*_{12,w}} = 0 \,, 
\label{h12eomShearI}
\end{equation}
are identical to those obtained from the original, uncomplexified effective action.
The crucial point of our analysis is that, on shell, the complexified effective action for the fluctuation can be re-expressed as a total derivative,
\begin{equation}
\call_{\complex}=16\pi G_N\left( h_{12,w}^*\cdot \frac{\dd S^{(2)}}{\dd h^*_{12,w}}\right)+\del_r J_w\,,
\label{totalLshearI}
\end{equation}
where the current $J_w$ is such that its imaginary part is radially conserved, as we already mentioned in (\ref{conscurrent}).
Further details regarding  $J_w$ will be given in the subsections below.

Using the standard holographic prescription, 
the retarded Green's function in \eqref{kuboeta} is obtained from the \emph{boundary} limit of the current $J_w$. Schematically, we have  
\begin{equation}\label{ImGR}
G_R(w)=\frac{1}{8\pi G_N}\lim_{r\to 0}\left(J_w+J_{GH}+J_{ct}\right),
\end{equation}
where $J_{GH}$ denotes the Gibbons--Hawking boundary term required for a well-defined variational principle \cite{Buchel:2004di},\cite{Cremonini:2009ih}, while $J_{ct}$ encodes holographic counterterms needed to regularize and renormalize the AdS boundary divergences \cite{Skenderis:2002wp}. Importantly, neither $J_{GH}$ nor $J_{ct}$ can contribute to the imaginary part of the Green's function, since these boundary terms are real even after complexification, see \cite{Buchel:2023fst}. 
Combining this with the fact that $\mathrm{Im}\,J_w$ is radially conserved (see the discussion by \eqref{conscurrent}) and may therefore be evaluated at the horizon, we have 
\begin{equation}
\Im G_R(w)=\frac{1}{8\pi G_N}\lim_{r\to 0}\Im J_w=\frac{1}{8\pi G_N}\lim_{r\to r_h}\Im J_w\,.
\label{retardedG}
\end{equation}
Finally, by decomposing the current into real and imaginary parts,
\begin{equation}\label{currentshearpieces}
    J_w=J_0-i\mathfrak{w}\,J_1\,, \quad\quad \mathfrak{w}=\frac{w}{2\pi T}\,,
\end{equation}
and combining \eqref{kuboeta} with \eqref{retardedG}, we find the following expression for the shear viscosity,
\begin{equation}\label{KuboShear}
    \eta= -\frac{1}{8\pi G_N}\lim_{w\to 0}\frac{1}{w}\,\Im J_w
    = \frac{1}{8\pi G_N}\cdot\frac{1}{2\pi T}\cdot J_1 \,,
\end{equation}
which will be evaluated at the horizon for each model below.

\subsubsection*{Model I}

Recall that Model I \eqref{action1} extends the setup of \cite{Buchel:2023fst} by including all possible four-derivative terms involving gravity and a $U(1)$ gauge field, and allowing the couplings of the higher-derivative terms which appear in \eqref{HDLag1} to 
be scalar dependent. 
For Model I, then,  
the complexified Lagrangian associated with \eqref{h12actionShearI} has the following form,
\begin{align}
   &\call_{\complex}=
   \frac{c_1}{c_2c_3}\left(h_{12,w}h_{12,-w}^{\prime\prime}+h_{12,-w}h''_{w}\right)
   +\Bigg(\frac{c_1'}{c_2c_3}-\frac{3c_1c_2'}{c_2^2c_3}-\frac{c_1c_3'}{c_2c_3^2}\Bigg)\biggl(h_{12,w}'h_{12,-w}
   \nonumber\\&+h_{12,w}h_{12,-w}^{\prime}\biggr)
   +\frac{3c_1}{2c_2c_3}h_{12,w}'h_{12,-w}^{\prime}+h_{12,w}h_{12,-w}\Bigg(\frac{\omega^2 c_3}{2c_1c_2}-\frac{6c_1c_3}{c_2}+\frac{c_1}{4c_2c_3}\sum_i\left(\phi_i'\right)^2\nonumber\\&+\frac{c_1c_3}{2c_2}V 
   -\frac{c_1'c_2'}{c_2^2c_3}+\frac{5c_1c_2'^2}{c_2^3c_3}-\frac{c_1'c_3'}{c_2c_3^2}+\frac{c_1c_2'c_3'}{c_2^2c_3^2}-\frac{\gamma A'^2}{4c_1c_2c_3}+\frac{c_1''}{c_2c_3}-\frac{c_1c_2''}{c_2^2c_3}\Bigg)+\mathcal{O}(\beta) \,, 
\end{align}
where $\mathcal{O}(\beta)$ contains contributions from the higher-derivative corrections.
The radial current $J_w$ we defined in \eqref{totalLshearI} is of the form
\begin{align}
J_{w}=&\Bigg[
B_1 h_{12,w}''+B_2h_{12,w}'
+\Big(B_3 w^2+A_1+B_4\Big)h_{12,w}
\Bigg]h_{12,-w}' +\Bigg[
- B_1 h_{12,w}'''-B_1' h_{12,w}''
\nonumber\\
&+\Big(B_5w^2+B_6
+\frac{1}{2}A_1\Big)h_{12,w}' 
+\Big(B_7w^2
+A_2
+B_{8}\Big)h_{12,w}
\Bigg]h_{12,-w}\,,\label{shearcurrent}
\end{align}
with the coefficients $A_i\sim \calo(\beta^0)$ and
$B_i\sim \calo(\beta)$ given explicitly in Appendix \ref{app:AppendixAI}. We have explicitly verified that the imaginary part of the current is conserved.

Next, to proceed, it is convenient to introduce the following replacement for the fluctuation,
\begin{align}\label{Hcapital}
    h_{12,w}=c^2_2\,H_{12,w}
    \,,\quad\quad
    H_{12,w}&=\left(\frac{c_1}{c_2}\right)^{-i\mathfrak{w}}\left(H_{0}+i\mathfrak{w}\,H_{1}\right)\,.
\end{align}
Using \eqref{Hcapital}, together with the incoming-wave boundary condition for the fluctuations and the appropriate boundary normalization, we have
\begin{equation}\label{boundnorm}
\begin{split}
\lim_{r\to 0} H_0(r)=1& \,, \qquad \lim_{r\to 0} H_1(r)=0 \,,\\
\lim_{r\to r_h} H_0(r)=\text{finite}& \,, \qquad \lim_{r\to r_h} H_1(r)=\text{finite} \,.
\end{split}
\end{equation}
Recalling the decomposition of the current in \eqref{currentshearpieces}, 
we can further separate the two- and four-derivative contributions to its imaginary part in the following way, 
\begin{equation}
    J_1\equiv F+\beta\cdot\delta F\,, 
\end{equation}
where the leading contribution is given by 
\begin{equation}\label{FShear}
    F=
    \frac{c_2^2\left(c_1c_2^{\prime}-c_2c_1^{\prime}\right)}{2c_3}H_{0}^2
    +
    \frac{c_1c_2^3}{2c_3}\left(H_{0}H_{1}^{\prime}
    -H_{0}^{\prime}H_{1}\right) \,,
\end{equation}
and is identical to the result of \cite{Buchel:2023fst}, while the higher-derivative corrections are 
\begin{align}\label{deltaFShear}
    \delta F=&\frac{1}{6c_1^2c_3^3}\Big(H_0^2\left(c_1c_2^{\prime}-c_2c_1^{\prime}\right)+c_1c_2\left(H_0H_1^{\prime}-H_1H_0^{\prime}\right)\Big) \nonumber\\
&\times\Bigg(- 12 c_1c_2(c_1'c_2+c_1c_2')(\sum_i\partial_i\alpha_3\cdot\phi'_i)
+4c_1\Big[3c_2c_1'\big(\left(3\alpha_1+3\alpha_2+8\alpha_3\right)c_2'
\big)  \nonumber\\
&
+c_1\Big(2c_2^2c_3^2\big(-12+V\big)\big(2\alpha_1+\alpha_2+2\alpha_3\big)
+ 9\big(\alpha_1+\alpha_2+3\alpha_3\big)\,c_2'^2
\Big)
\Big] \nonumber \\
&
+ c_2^2\Big(
\gamma\left(2\alpha_1+\alpha_2+2\alpha_3\right)
-3\left(2\alpha_4+\alpha_5\right)
\Big)A'^2\Bigg) \,.
\end{align}
By setting the scalar-dependent couplings to constants and the gauge-field contributions to zero, we recover the corresponding expression in \cite{Buchel:2023fst}.

To compute the shear viscosity in \eqref{KuboShear}, we evaluate at the horizon the imaginary parts of the current, \eqref{FShear} and \eqref{deltaFShear}, using \eqref{BGfunctions1} and \eqref{phiHRZN}. We find (see \eqref{defco} for the
definition of $\calc$)
\begin{align}
    F\sim & \left(4G_Ns_B\right)\left(2\pi T\right)\cdot\left(\frac{1}{2}H_0^2\right)\,,\label{Fhor}\\
    \delta F\sim & \left(4G_Ns_B\right)\left(2\pi T\right)\cdot \frac{H_0^2}{3}\Bigg[(V-12)(5\alpha_1+\alpha_2)-3
    (\sum_i\partial_iV\cdot\partial_i\alpha_3)\label{dFhor}\\
    &- \frac{\calc}{2}\Big(\gamma\big(\alpha_1+2\alpha_2+6\alpha_3\big)+3\big(2\alpha_4+\alpha_5-(\sum_i\partial_i\gamma\cdot\partial_i\alpha_3)\big)\Big)\Bigg] \,. \nonumber
\end{align}
Once we turn off the contributions associated with the presence of the gauge field, the result matches  \cite{Buchel:2023fst}, with the leading-order contribution \eqref{Fhor} unchanged.

To proceed with the evaluation of \eqref{Fhor}-\eqref{dFhor}, we solve the fluctuation equations of motion derived from \eqref{h12eomShearI} after implementing the replacements in \eqref{replacemnt1} and \eqref{Hcapital}. Since \eqref{Fhor}-\eqref{dFhor} depend only on the value of $H_0$ at the black-brane horizon, it is sufficient to determine the solution for the $H_0$ mode.
In the hydrodynamic limit $\omega =0$, the equation of motion for $H_0$ takes the form
\begin{equation}
    0 = H_0'' + H_0'\left[\ln\,\frac{c_1 c_2^{3}}{c_3}\right]' + \beta \cdot \delta eq \,,
\end{equation}
where $\delta eq $ contains contributions from higher-derivative terms.
To solve this equation, we are looking for solutions recursively in $\beta$,
\begin{equation}
    H_0=H_{0,0}+\beta\,H_{0,1}\,,
\end{equation}
where $H_{0,0}$ is the leading contribution 
\begin{equation}\label{H00}
    H_{0,0}=\mathcal{C}_{1,0}+\mathcal{C}_{2,0}\int dr\,\frac{c_3}{c_2^3c_1} \,,
\end{equation}
and $H_{0,1}$ the correction.
Demanding regularity of $H_{0,0}$ at the horizon sets $\mathcal{C}_{2,0}=0$, 
while the boundary normalization as $r\to 0$ fixes $\mathcal{C}_{1,0}=1$.
Since $H_{0,0}\equiv 1$, the corresponding expression for $H_{0,1}$ takes the form of \eqref{H00}. The boundary conditions \eqref{boundnorm}, however, force the integration constants to vanish, $\mathcal{C}_{1,1}=\mathcal{C}_{2,1}=0$. 
Therefore, collecting everything, we find
\begin{equation}\label{H01}
    H_0(r) \equiv 1 + \mathcal{O}(\beta^2)\,.
\end{equation}
Finally, using \eqref{Fhor}-\eqref{dFhor} and \eqref{H01} to evaluate
\begin{equation}
\label{mod1etaa}
    \eta
    = \frac{1}{8\pi G_N}\cdot\frac{1}{2\pi T}\cdot\left(F+\beta\,\delta F\right)
\end{equation}
at the horizon gives the shear viscosity-to-entropy ratio 
reported in \eqref{etasmod1}. Note that \eqref{mod1etaa} naturally
gives the ratio $\frac{\eta}{s_B}$. Thus, to obtain the physical ratio
$\frac{\eta}{s}$, the latter must be corrected by the ratio
$\frac{s_B}{s}\equiv 1-\beta\cdot \kappa$,  as given by \eqref{swsb}
and \eqref{sWsBRatio}.

\subsubsection*{Model II}
Parallel to the discussion for Model I, we start the analysis by evaluating the action \eqref{action2} on the background with fluctuations \eqref{h12}. Once we write fluctuation in the form of \eqref{replacemnt1}, the complexified action can be rewritten as in \eqref{totalLshearI}, from which we can obtain the current $J_w$. To first order in $\epsilon$ and fully nonlinear in
$\lambda_1$, the current for this model is given by
\begin{align} \label{ShearJModelTwo}
       & J_w = \, \bigg[ -\frac{2 c_1 c_2'}{c_2^2 c_3} + \left( \frac{2 w^2 c_1'}{c_1^2 c_2 c_3} + \frac{4 w^2 c_2'}{c_1 c_2^2 c_3} + \frac{12 c_1' c_2'^2}{c_2^3 c_3^3} + \frac{4 c_1 c_2'^3}{c_2^4 c_3^3}  \right) \, \lambda_1  \nonumber \\
        &+ \left(\frac{2 w^2}{c_1 c_2 c_3}  + \frac{4 c_1' c_2'}{c_2^2 c_3^3}\right) \, \sum_i \partial_i \lambda_1 \cdot \phi_i'  \bigg] \, h_{12,w} h_{12,-w} + \bigg[ \frac{c_1}{2 c_2 c_3} - \left( \frac{2 w^2}{c_1 c_2 c_3} + \frac{2 c_1' c_2'}{c_2^2 c_3^3}\right) \, \lambda_1 \nonumber \\
        &+\left( \frac{2 c_1' }{c_2 c_3^3} + \frac{2 c_1 c_2'}{c_2^2 c_3^3}\right) \, \sum_i \partial_i \lambda_1 \cdot \phi_i' \bigg] \, h_{12,w}' h_{12,-w} + \bigg[ \frac{c_1}{c_2 c_3} - \left( \frac{2 w^2}{c_1 c_2 c_3} + \frac{4 c_1' c_2'}{c_2^2 c_3^3}\right) \, \lambda_1 \bigg]  \nonumber \\
        &\times \, h_{12,w} h_{12,-w}' - \bigg[ \left(\frac{2 c_1'}{c_2 c_3^3} + \frac{2 c_1 c_2'}{c_2^2 c_3^3} \right) \, \lambda_1 \bigg] \, h_{12,w}' h_{12,-w}'  + \epsilon  \cdot  \Big[ \ldots  \Big] \, .
\end{align}
We note that only $h_{12,w}$ and its first derivative are present in the current above, and therefore we do not need to directly solve and use the equations of motion for $h_{12,w}$ in order to further evaluate the current. Meanwhile, we may use the background equations of motion for the gauge field \eqref{GBBGEOMS1}-\eqref{GBBGEOMS2} to eliminate $A''$ which appears in the $\mathcal{O}(\epsilon)$ part of \eqref{ShearJModelTwo}.

As before, the imaginary part of the current \eqref{ShearJModelTwo} is conserved, and thus can be used to calculate the retarded correlation function following \eqref{kuboeta}. In the hydrodynamic limit, and using the same variables $H_0$ and $H_1$ as in \eqref{Hcapital} with boundary normalization \eqref{boundnorm}, we find  the current  
\begin{align}
        J_w = \, &J_0 - i \mathfrak{w} \, J_1 \, , \\
        J_1 = \, &\frac{H_0^2}{2 c_3}\, (c_2^3 c_1' - c_1 c_2^2 c_2') + \frac{c_1 c_2^3 }{2 c_3} (H_1 H_0' - H_0 H_1')- \frac{1}{c_1 c_3^3} \,(2 c_2^2 c_1'^2 c_2' H_0^2 - 2 c_1 c_2 c_1' c_2'^2 H_0^2 \nonumber \\
        &+ 2 c_1 c_2^2 c_1' c_2' H_1 H_0' - 2 c_1 c_2^2 c_1' c_2' H_0 H_1') \, \lambda_1 + 
        \frac{1}{c_1 c_3^3} (2 c_1^2 c_2 c_2'^2 H_0^2 - 2 c_2^3 c_1'^2 H_0^2 \nonumber \\ 
        &- 2 c_1 c_2^3 c_1' H_1 H_0'-2 c_1^2 c_2^2 c_2' H_1 H_0' +  2 c_1 c_2^3 c_1' H_0 H_1' + 2 c_1^2 c_2^2 c_2' H_0 H_1') \, \sum_i \partial_i \lambda_1 \cdot \phi_i' \nonumber\\
        &+ \epsilon  \cdot  \Big[ \ldots  \Big] \, ,
\end{align}
from which we can calculate the shear viscosity using \eqref{KuboShear}.

Since $J_1$ is radially conserved, we can evaluate it at $r \rightarrow r_h$, which simplifies the expression significantly. The result,
besides the horizon values of the couplings $\lambda_i$ and scalar
potential $V$, and their
derivatives with respect to the bulk scalars
$\frac{\del_j\lambda_i}{\del\phi_j}$ and $\frac{\del_j V}{\del\phi_j}$,
depends on the horizon value of $H_0$,
but is independent of $H_1$ --- much like in the equations \eqref{Fhor} and
\eqref{dFhor} of Model I. It is straightforward to
check (using the background equations of motion for the Model II)
that the solution of the equation of motion for $H_0$ obtained from 
 from \eqref{h12eomShearI}, after implementing the replacements in \eqref{replacemnt1} and \eqref{Hcapital}, in the hydrodynamic limit $w=0$, is solved by
 (compare with \eqref{H01})
 \begin{equation}
 H_0(r)\equiv 1+\calo(\epsilon^2) \, .
 \label{h01mod2}
 \end{equation}
Using the result \eqref{h01mod2} we arrive, following identical steps
as in Model I, to \eqref{etasmod2}. Note that in this model
there is no difference between the Bekenstein entropy and the Wald
entropy of the black brane \eqref{sWsBModelTwo}.  
Finally, the usual Gauss-Bonnet theory can be recovered by setting $\epsilon =0$, 
which agrees with \cite{Buchel:2024umq} when we turn off the gauge field and the scalar-dependence of $\lambda_1$.

\subsection{Bulk Viscosity}\label{sec:bulkviscosity}

The starting point for the calculation of the bulk viscosity is the Kubo formula \cite{Gubser:2008sz},
\begin{equation} \label{BulkKubo}
    \zeta = - \frac{4}{9} \lim_{w \rightarrow 0} \frac{1}{w} \operatorname{Im} G_R(w), \quad G_R(w) = - i \int dt d\boldsymbol{x} e^{i w t} \theta(t) \langle
[\frac{1}{2}T_i^{i}(t,\boldsymbol{x}),\frac{1}{2}T_j^j(0,\boldsymbol{0})]\rangle\,.
\end{equation}
We compute the relevant retarded correlation function for a decoupled set of $SO(3)$ invariant fluctuations for the metric, gauge field and bulk scalars
\cite{Buchel:2010gd}
given by 
\begin{equation}
    \begin{split}
        &ds_5^2 \rightarrow ds_5^2 + h_{tt}(t,r) dt^2 + h_{11}(t,r) d\boldsymbol{x}^2 + 2 h_{tr}(t,r) dt dr + h_{rr} dr^2,\\
        &A_\mu \rightarrow A_\mu + \Big(a_t(t,r),0,0,0, a_r(t,r) \Big), \quad\quad\quad \phi_i \rightarrow \phi_i + \psi_i(t,r)\,.
    \end{split}
\end{equation}
As for gauge fixing, we use the freedom we have to set the following fluctuations to zero\footnote{We impose this condition only after we derived the EoMs from the effective action.},
\begin{equation}
\label{zetagauge}
    h_{tr} = h_{rr} = a_r =0.
\end{equation}
Moreover, for convenience we introduce new variables  for the fluctuations, 
\begin{equation} \label{BulkHdef1}
    \begin{split}
    h_{tt}(t,r) = e^{-iwt} c_1^2 \, H_{00}(r), \quad\quad h_{11}(t,r) = e^{-iwt} c_2^2 \, H_{11}(r)\,,
    \end{split}
\end{equation}
together with a set of gauge invariant fluctuations $Z_i$ and $Z_a$ as in \cite{Benincasa:2005iv,Buchel:2010gd},
\begin{equation} \label{BulkHdef2}
    \psi_i (t,r) = e^{-iwt} \Big( Z_i(r) + \frac{\phi_i'c_2}{2 c_2'} H_{11} \Big)\,, \quad\quad a_t(t,r) = e^{-iwt} \Big( Z_a(r) + \frac{A'c_2}{2 c_2'} H_{11} \Big)\,.
\end{equation}
Using the above gauge invariant fluctuations, the linear coupled
system of the second-order  scalar equations $\{Z_i\}$
completely decouples from the remaining fluctuations.
The coupled linear system of the first-order equations for 
$\{H_{00},H_{11},Z_a\}$ (coming from the gauge fixing  \eqref{zetagauge}
constraints) depends on the solution for the set of the scalar
fluctuations $\{Z_i\}$.

Similar to the analysis for the shear viscosity, we start by computing the complexified action for the fluctuations. Writing the fluctuations as
\begin{equation}
    h_{tt}(t,r) = e^{-iwt} h_{00,w}\,,\,  h_{11}(t,r) = e^{-iwt} h_{11,w}\,,\,  \psi_i(t,r) = e^{-iwt} \varphi_{i,w}\,,\, a_t(t,r) = e^{-iwt} a_{w}\,,
\end{equation}
the action can be written schematically as a functional of $\lbrace h_{00,w},h_{11,w},\varphi_{i,w},a_{w} \rbrace$ and their complex conjugates,
\begin{equation}
    S^{(2)} = \frac{1}{16 \pi G_N} \int_0^{r_h} dr \; \mathcal{L}_{\mathbb{C}}\lbrace h_{00,w},h_{11,w},\varphi_{i,w},a_{w},h_{00,w}^\ast,h_{11,w}^\ast,\varphi_{i,w}^\ast,a_{w}^\ast \rbrace \,.
\end{equation}
As before, we can extract the current by rewriting the on-shell effective action as
\begin{equation}
    \mathcal{L}_{\mathbb{C}} = 16 \pi G_N \cdot \Big( h_{00,w}^\ast \cdot \frac{\delta S^{(2)}}{\delta h_{00,w}^\ast} +  h_{11,w}^\ast \cdot \frac{\delta S^{(2)}}{\delta h_{11,w}^\ast} + \varphi_{i,w}^\ast \cdot \frac{\delta S^{(2)}}{\delta \varphi_{i,w}^\ast} +  a_{w}^\ast \cdot \frac{\delta S^{(2)}}{\delta a_{w}^\ast} \Big) + \partial_r J_w\,,
\end{equation}
where $J_w$ is the current which we later verify to be radially conserved.

 \subsubsection*{Model I}
Following the procedure described above, we find that the current $J_w$ for Model I is
\begin{align} \label{BulkCurrent}
         J_w = \; &\frac{c_2^3 \gamma A'}{4 c_1^3 c_3} \,  h_{00,w}a_{-w} + \frac{3 c_2 \gamma A'}{4 c_1 c_3} \, h_{11,w} a_{-w} + \frac{c_2^3 \gamma}{2 c_1 c_3}  \, a_{w}'a_{-w} - \frac{3 c_2^3 c_1'}{4 c_1^4 c_3} \, h_{00,w}h_{00,-w} \nonumber \\ 
         &- \frac{3 c_2 c_1'}{4 c_1^2 c_3} \,  h_{11,w}h_{00,-w} - \frac{3 c_2'}{4 c_1 c_3} \, h_{00,w}h_{11,-w} -  \frac{3 c_1 c_2'}{4 c_2^2 c_3} \,h_{11,w}h_{11,-w} + \frac{c_2^3}{4 c_1^3 c_3} \, h_{00,w}h_{00,-w}' \nonumber \\
         &+ \frac{3 c_2}{4 c_1 c_3} \, h_{11,w}h_{00,-w}'+ \frac{c_2^3}{4 c_1^3 c_3} \, h_{00,w}'h_{00,-w}+ \frac{3 c_2}{4 c_1 c_3} \, h_{00,w}h_{11,-w}'- \frac{3 c_1}{4 c_2 c_3} \, h_{11,w}h_{11,-w}' \nonumber \\
         &+ \frac{3 c_1}{4 c_2 c_3} \, h_{11,w}'h_{11,-w} - \frac{c_1 c_2^3}{2 c_3} \, \sum_i \varphi_{i,w}' \cdot \varphi_{i,-w} + \frac{c_2^3 A'}{2 c_1 c_3} \,  \sum_i (\partial_i \gamma) \cdot \varphi_{i,w} a_{-w} \nonumber \\
         &+ \frac{c_2^3}{4 c_1 c_3} \,  \sum_i \phi_i' \cdot  h_{00,w}\varphi_{i,-w}   - \frac{3 c_1 c_2}{4 c_3}  \, \sum_i \phi_i' \cdot h_{11,w} \varphi_{i,-w}    + \beta  \cdot  \Big[ \ldots  \Big]\,.
\end{align}
In terms of the new variables \eqref{BulkHdef1}-\eqref{BulkHdef2}, the equations of motion for the fluctuations in this model can be written as
\begin{align} \label{BulkEq2}
    &Z_i'' + \left( \ln{\frac{c_1 c_2^3}{c_3}}\right)' Z_i'+ \frac{w^2 c_3^2}{c_1^2}\, Z_i
 + (12 - V) \,\frac{ c_2^2 c_3^2}{9 c_2'^2} \, (\phi_i') \, \sum_j Z_j \cdot \phi_j' - c_3^2 \sum_j Z_j \cdot (\partial_{ij}^2 V) \nonumber 
\\
    &- \frac{c_2 c_3^2}{3 c_2'} \left( (\partial_i V) \sum_j Z_j \cdot \phi_j' + (\phi_i') \sum_j Z_j \cdot (\partial_j V) \right)- \frac{A'^2}{c_1^2 \gamma} \, (\partial_i \gamma) \,\sum_j Z_j \cdot (\partial_j \gamma) 
\nonumber \\
    &+ \frac{c_2 A'^2}{6 c_1^2 c_2'} \left( (\partial_i \gamma) \sum_j Z_j \cdot \phi_j' + (\phi_i') \sum_j Z_j \cdot (\partial_j \gamma) \right) + \frac{A'^2}{2 c_1^2}\sum_j Z_j \cdot (\partial_{ij}^2 \gamma)
\nonumber \\
    & -\frac{c_2^2 \gamma A'^2}{18 c_1^2 c_2'^2} \, (\phi_i') \, \sum_j Z_j \cdot \phi_j'  + \beta  \cdot  \Big[ \ldots  \Big] =0 \, , 
\end{align}
and 
\begin{equation} \label{BulkEq1}
\begin{split}
     &H_{11} \equiv  \frac{c_2'}{c_2 c_3} H , \quad H' + \frac{c_2 c_3}{3 c_2'} \sum_i Z_i \cdot \phi_i' + \beta  \cdot  \Big[ \ldots  \Big] =0\,,
     \end{split}
\end{equation}
\begin{align} \label{BulkEq4}
     &H_{00}' + \left(\frac{1}{3} c_3 (12 - V) - \frac{w^2 c_3}{c_1^2}  + \frac{\gamma A'^2}{3 c_1^2 c_3} - \frac{c_1'^2}{c_1^2 c_3} - \frac{3c_1' c_2'}{c_1 c_2 c_3} \right) H - \frac{c_2 c_3^2}{3 c_2'} \, \sum_i Z_i \cdot (\partial_i V)  \nonumber \\
    &\quad\,\,\, + \frac{c_2 A'^2}{6 c_1^2 c_2'} \, \sum_i Z_i \cdot (\partial_i \gamma) + \left( \frac{c_2^2 c_3^2}{9 c_2'^2} \, (12 - V) - \frac{c_2 c_1'}{3 c_1 c_2'} - \frac{c_2^2 \gamma A'^2}{18 c_1^2 c_2'^2} \right) \, \sum_i Z_i \cdot \phi_i'   \nonumber \\
    &\quad\,\,\, + \frac{c_2}{3 c_2'} \, \sum_i Z_i' \cdot \phi_i'  + \beta  \cdot  \Big[ \ldots  \Big] =0\,,
\end{align}
\begin{equation} \label{BulkEq5}
\begin{split}
    &Z_a' + \frac{A'}{2}  H_{00} + \frac{c_1' A'}{2 c_1 c_3} \, H + \frac{A'}{\gamma}\, \sum_i Z_i \cdot (\partial_i \gamma) - \frac{c_2 A'}{6 c_2'} \, \sum_i Z_i \cdot \phi_i'  + \beta  \cdot  \Big[ \ldots  \Big] =0 \, .
\end{split}
\end{equation}

We now need to solve \eqref{BulkEq2}-\eqref{BulkEq5} perturbatively in $\beta$ and in the hydrodynamic approximation, \textit{i.e.} to the first order in $w$. Moreover, to properly impose the boundary conditions, it is convenient to use the following new variables from now on
\begin{equation} \label{BulkCapVar}
\begin{split}
     &Z_i = \left( \frac{c_1}{c_2}\right)^{-i\mathfrak{w}} (z_{i,0} + i \mathfrak{w} \, z_{i,1}) \,, \quad Z_a = \left( \frac{c_1}{c_2}\right)^{-i\mathfrak{w}} (z_{a,0} + i \mathfrak{w}  \, z_{a,1})\, ,\\ &H = H_0 + i \mathfrak{w}  \, H_1 \,, \quad H_{00} = H_{00,0} + i \mathfrak{w}  \, H_{00,1} \,,
\end{split}
\end{equation}
where $\mathfrak{w} = w/(2\pi T)$, as before.
The incoming wave boundary condition implies that $z_{i,0},z_{i,1}$,$z_{a,0}$, and $z_{a,1}$ must be regular at the black brane horizon. Moreover, to correctly normalize the retarded correlation function, we impose the following conditions at the boundary, 
\begin{equation}
    \lim_{r \rightarrow 0} H_{11}(r) =1\,, \qquad  \lim_{r \rightarrow 0} H_{00}(r) =0\,, \qquad \lim_{r \rightarrow 0} Z_a(r) =0 \,.
\end{equation}
Additionally, the coefficients  of $\psi_i$ that are non-normalizable near the AdS boundary
must vanish \cite{Benincasa:2005iv}.
From \eqref{BulkCapVar} this implies that if the non-normalizable coefficient $\lambda_i$ of the background scalar
$\phi_i$ dual to a gauge theory operator of dimension $\Delta_i$ is nonzero, \ie 
\begin{equation}
\phi_i=\lambda_i\cdot r^{4-\Delta_i}+\cdots\,,\qquad {\rm as}\  r\to 0 \,,
\label{nonnorm}
\end{equation}
the near-AdS boundary asymptotic of the corresponding $Z_i$ much be
\begin{equation}
\lim_{r\to 0}\ \frac{Z_i}{r^{4-\Delta_i}}=\frac {4-\Delta_i}{2}\cdot \lambda_i=\lim_{r\to 0}\ \frac{z_{i,0}}{r^{4-\Delta_i}}
\,,\qquad \lim_{r\to 0}\ \frac{z_{i,1}}{r^{4-\Delta_i}}=0\,.
\label{zibads}
\end{equation}

Using equations \eqref{BulkEq2}-\eqref{BulkEq5}, we can verify that the imaginary part of the current
\eqref{BulkCurrent} is indeed radially conserved and thus, following our previous discussion, we can use it to calculate the retarded correlation  function at the horizon.
To further evaluate the current, we use the equations for the fluctuations
\eqref{BulkEq1}-\eqref{BulkEq5} to eliminate all derivatives of $H_{11}$, $H_{00}$ and $Z_a$.
Together with \eqref{BulkEq2} and the background equations \eqref{BGEOMS3}-\eqref{BGEOMS1}, we
find\footnote{While we will not present here the result for $\dd F$, we note that much like $F$
\eqref{BulkJF},
it is a functional of $z_{i,0}$ and  $z_{i,1}$ only as well.}
\begin{equation}
    \begin{split} \label{BulkJF}
        &J_w = J_0 - i \mathfrak{w} \, J_1, \quad J_1 \equiv F + \beta \cdot \delta F \,,\\
        &F = - \frac{c_2^2 (c_2 c_1' - c_2' c_1)}{2 c_3} \sum_i (z_{i,0})^2 + \frac{c_2^3 c_1}{2 c_3} \sum_i (z_{i,1}' z_{i,0} - z_{i,0}' z_{i,1}) \,,
    \end{split}
\end{equation}
from which we can calculate the bulk viscosity as (see \eqref{BulkKubo} and \eqref{BulkJF})
\begin{equation}
    \zeta = - \frac{4}{9} \cdot \frac{1}{8 \pi G_N}  \lim_{w \rightarrow 0}  \frac{1}{w} \operatorname{Im} J_w=  \frac{4}{9} \cdot \frac{1}{8 \pi G_N}  \lim_{w \rightarrow 0} \frac{\mathfrak{w}}{w} J_1 =  \frac{4}{9} \cdot \frac{1}{8 \pi G_N} \cdot \frac{1}{2 \pi T} \cdot (F + \beta \cdot \delta F).
\end{equation}
Evaluating this at the horizon, and using 
\eqref{BGfunctions1}-\eqref{phiHRZN} together with \eqref{Temp}, \eqref{Bekenstein} and \eqref{sWsBRatio},  we eventually find the expression for the ratio $\frac{\zeta}{s}$ reported in \eqref{zetasmod1}.

For the case of a charged plasma, the framework presented here has not been verified with other approaches yet\footnote{It has been extensively validated
for the neutral plasma in \cite{Buchel:2023fst}.}. 
The first computation of the bulk viscosity in
a two-derivative holographic model, albeit with perturbative breaking
of the conformal invariance, was performed in \cite{Buchel:2010gd}.
That result was later verified in \cite{Eling:2011ms}. In Appendix \ref{app:AppendixB}
we reproduced the computation in \cite{Buchel:2010gd} following our discussion
here.

\subsubsection*{Model II}

Computation of the bulk viscosity in Model II \eqref{defmod2f}
proceeds as in Model I (see the previous section). While the final result 
is remarkably simple \eqref{zetasmod2}, the technical details
to reach it are extremely complicated. Here we overview the
general structure of the relevant equations only. 

To obtain the bulk viscosity in Model II, one needs to solve
the decoupled set of equations of motion for the gauge
invariant bulk scalar fluctuations  $\{z_{i,0}\}$, see \eqref{BulkHdef2}
and \eqref{BulkCapVar}, in the hydrodynamic
$\mathfrak{w}=0$ limit, subject to regularity at the black brane horizon
and the asymptotic $AdS_5$ boundary expansion as in \eqref{zibads}.
The ratio of the bulk viscosity to the entropy density
\eqref{zetasmod2} is determined from the black brane horizon
values of these scalars, to order $\calo(\epsilon)$. 
Unfortunately, the equations of motion for $z_{i,0}$ are rather involved\footnote{
For a bulk field independent $\lambda_1$ and with $A=0$
their explicit form is given in \cite{Buchel:2024umq}.},
\begin{equation}
\begin{split}
&0=z_{i,0}''+\left(\ln\frac{c_1 c_2^3}{c_3}\right)'\, z_{i,0}'+\biggl(\calw_1\cdot\phi_i'+\calw_2\cdot\del_i\lambda_1\biggr)\cdot \left(\sum_{jk}z_{k,0}\cdot \phi_j''\cdot \del^2_{kj}\lambda_1\right) \\
&-\calw_3\cdot \left(\sum_j z_{j,0}\cdot \del^2_{ij}\lambda_1\right)-c_3^2\cdot \left(\sum_j z_{j,0}\cdot \del^2_{ij}V\right)
+\frac{(A')^2}{2 c_1^2}\cdot \left(\sum_j z_{j,0}\cdot \del^2_{ij}\gamma\right)  \\
&-\frac{(A')^2}{c_1^2}\cdot \left(\sum_j z_{j,0}\cdot \del_j\gamma\right)\cdot \del_i\ln \gamma+\left(\sum_{j}\phi_j''\cdot \del_j\lambda_1\right)
\cdot \biggl(\calt_1\cdot \phi_i'+\calt_2\cdot \del_i\lambda_1\biggr)\\
&+\calt_3\cdot \phi_i''+\left(\sum_j(\phi_j)^2\right)\cdot \calt_4\cdot \phi_i'
+\calt_5\cdot \del_i\lambda_1+\calt_6\cdot \phi_i'+\epsilon\cdot \biggl[\cdots\biggr]\, ,
\end{split}
\label{zi0}
\end{equation}
where $\calt_a$, $a=1\cdots 6$ are linear combinations of 
\begin{align}
&\sum_i z_{i,0}'\cdot \del_i\lambda_1\,,\quad \sum_i z_{i,0}\cdot \del_i\lambda_1\,,\quad \sum_i z_{i,0}'\cdot \phi_i'\,,\quad \sum_i z_{i,0}\cdot \phi_i'\,,\quad \sum_{ij}z_{i,0}'\cdot \phi_j'\cdot \del^2_{ij}\lambda_1 \, ,  \nonumber \\
&\sum_{ij}z_{i,0}\cdot \phi_j'\cdot \del^2_{ij}\lambda_1\,,\quad
\sum_{ij} z_{i,0}\cdot \del_j\lambda_1\cdot \del^2_{ij}V\,,\quad \sum_{ijk}z_{i,0}\cdot \phi_j'\cdot \phi_k'
\cdot\del^3_{ijk}\lambda_1 \, ,\\
&\sum_{ij} z_{i,0}\cdot \del_j\lambda_1\cdot \del^2_{ij}\lambda_1\,,\quad
\sum_{ij} z_{i,0}\cdot \del_j\lambda_1\cdot \del^2_{ij}\gamma\,,\quad \sum_{i} z_{i,0}\cdot \del_i V
\,,\quad \sum_{i} z_{i,0}\cdot \del_i \gamma \, ,  \nonumber
\label{zstructures}
\end{align}
with coefficients, which, besides the background $\{c_1,c_2,c_3,\phi_i,A\}$ dependence, are rational
functionals of 
\begin{equation}
\begin{split}
&\sum_{i}\phi_i'\cdot \del_i\lambda_1\,,\quad
\sum_{i}(\del_i\lambda_1)^2\,,\quad
\sum_{i}\del_i\lambda_1
\cdot\del_i \gamma \, .
\end{split}
\label{cstructures}
\end{equation}
The numerators of these functionals are at most third-order polynomials of \eqref{cstructures},
and the denominators are (in general) a product of,  at most, second-order polynomials of \eqref{cstructures}
(of the total degree 4 at the most). The functionals $\calw_b$, $b=1\cdots 3$ are similar rational functions, with linear
in \eqref{cstructures} numerators, and (at most) a degree-2 in  \eqref{cstructures} polynomial denominators.
The terms in $[\cdots]$ of \eqref{zi0} are linear in $\lambda_2$ and its
derivatives.


\subsection{Conductivity}\label{sec:conductivity}

We are now ready to move on
 to the computation of the conductivity $\sigma$, for which Kubo's formula  is written in terms of the retarded two-point correlation function of the conserved boundary current $J_1$,
\begin{equation}\label{KuboConductivity}
\sigma = - \lim_{\omega \to 0} \frac{e^2}{\omega} \mathrm{Im}\,G_R(\omega)\, ,     \quad\quad
G_R(\omega)
= -i \int dt d\mathbf{x}\,e^{i\omega t}\theta(t)
\langle [J_1(t,\mathbf{x}), J_1(0,0)] \rangle  \,,
\end{equation}
where the index $1$ denotes the spatial direction $x_1$.
Here $e$ is a small coupling that gauges the global $U(1)$ symmetry of the boundary gauge theory, see \cite{Kovtun:2008kx}
for further discussion. 
We start by adding the following perturbations of the spacetime metric 
and the gauge field in \eqref{metric},
\begin{align}
    ds^2_5&\,\rightarrow\, 
    ds^2_5+2h_{tx}(t,r)dtdx_1
    +2h_{xr}(t,r)dx_1dr \,, \label{fluct1}\\
    A_{\mu}&\,\rightarrow\, 
    A_{\mu}
    +a(t,r)\, \delta^{x_1}_\mu \,.\label{fluct2}
\end{align}
Next, following the same steps explained in the shear and the bulk viscosity subsections, we complexify the fluctuations relevant for computing the conductivity.
We introduce harmonic time dependence for the gauge field fluctuation,
\begin{equation}
    a(t,r)=e^{-iwt}a_{w}(r) \,,\quad\quad
    a^{*}(t,r)=e^{iwt}a_{-w}(r) \,,
\end{equation}
and analogously for the metric perturbations $h_{tr}(t,r)$ and $h_{xr}(t,r)$,
such that the quadratic effective action for the fluctuations takes the form 
\begin{equation}\label{conductivityaction1}
S^{(2)}=\frac{1}{16\pi G_N}\int_0^{r_h} dr\,\mathcal{L}_{\mathbb{C}}\left\{h_{tx,w},h_{tx,w}^*,h_{xr,w},h_{xr,w}^*,a_w,a_{w}^*\right\} \,.
\end{equation}
Following \cite{Myers:2009ij}, we fix the gauge\footnote{It should be mentioned that the equations of motion for the fluctuations should be derived from the action \eqref{conductivityaction1} before fixing the gauge.} by setting
\begin{equation}
    h_{xr,w}= h_{xr,-w}\equiv 0 \,.
\end{equation}
The resulting complexified action, on-shell, can then be expressed as a total derivative, 
\begin{equation}\label{ConComplexL}
    \mathcal{L}_{\mathbb{C}}=
    16\pi G_N\left(h_{xt,w}^*\frac{\delta S^{(2)}}{\delta h_{xt,w}^*}
    +a_{w}^*\frac{\delta S^{(2)}}{\delta a_{w}^*}\right)+\partial_rJ_{w} \,,
\end{equation}
where the current $J_{w}$ contains contributions from all the fluctuations. Next, we decompose the current in real and imaginary parts,
\begin{equation}\label{currentcond}
    J_w=J_0-i\mathfrak{w}\,J_1\,,
\end{equation}
and from \eqref{ImGR} and \eqref{KuboConductivity} we find the following expression for the conductivity,
\begin{equation}\label{kubosigma}
\hat\sigma = -\lim_{\omega\to 0}\frac{e^2}{\omega}\,\Im J_\omega
    = e^2\lim_{\omega\to 0}\,\frac{\mathfrak{w}}{\omega}\,J_1 \,,
\end{equation}
which will be evaluated at the horizon for each model below.
Note we introduced here a conductivity-per-degree-of-freedom as
\begin{equation}
\hat\sigma\, \equiv\, 8\pi G_N\, \sigma\, \sim\, \frac{\sigma}{N^2} \, .
\label{defhs}
\end{equation}

\subsubsection*{Model I}

The gauge-fixed complexified Lagrangian for Model I is of the form
\begin{align}
\mathcal{L}_{\mathbb{C}}=&
-\frac{c_2}{c_1c_3}\left(h_{tx,w}h_{tx,-w}''+h_{tx,w}''h_{tx,-w}\right)
-\frac{c_1c_2\gamma}{2c_3}a_{w}'a_{-w}'
-\frac{3c_2}{2c_1c_3}h_{tx,w}'h_{tx,-w}'\nonumber \\
&+\left(
\frac{2c_2c_1'}{c_1^2c_3}
+\frac{c_2c_3'}{c_1c_3^2}
\right)\left(h_{tx,w}'h_{tx,-w}
+h_{tx,w}h_{tx,-w}'\right)
-\left(h_{tx,-w}a_w'+h_{tx,w}a_{-w}'\right)\frac{c_2\gamma A'}{2c_1c_3}\nonumber
\\
&+\frac{h_{tx,w}h_{tx,-w}}{c_1}\Bigg(
6c_2c_3-\frac{c_2\sum_i\left(\phi_i^{\prime}\right)^2}{4c_3}
-\frac{c_2c_3V}{2}
-\frac{2c_2c_1'^2}{c_1^2c_3}
-\frac{3c_1'c_2'}{c_1c_3}
+\frac{c_2'^2}{c_2c_3}
+\frac{c_2\gamma A'^2}{4c_1^2c_3}
\Bigg)\nonumber
\\
&
+\frac{\omega^2c_2c_3\gamma}{2c_1}a_{w}a_{-w}+\mathcal{O}(\beta)\,, 
\end{align}
where we have used the bulk equations of motion in \eqref{BGEOMS3} to replace $c_1^{\prime\prime}$ and $c_2^{\prime\prime}$.
The associated current 
(extracted from \eqref{ConComplexL}) is then given by
\begin{equation}\label{currentcondI}
\begin{split}
J_{w}=&\Bigg[
B_{9}h_{tx,w}''
+B_{10}h_{tx,w}'
+B_{11}a_w'
-\Big(\frac{c_2}{c_1c_3}
-B_{12}\Big)h_{tx,w}\Bigg]h_{tx,-w}'+\Bigg[B_{13}h_{tx,w}^{\prime\prime\prime} \\
&
+\Big(-\frac{c_2}{2c_1c_3}
+\omega^2B_{14}+B_{15}\Big)h_{tx,w}'
-B_{16}h_{tx,w}''
+B_{17}a_w^{\prime}
+\omega^2B_{18}a_{w}
\\
&
+h_{tx,w}\Big(\frac{c_2c_1^{\prime}}{c_1^2c_3}
+\frac{c_2^{\prime}}{c_1c_3}-\omega^2B_{19}+B_{20}\Big)\Bigg]h_{tx,-w}+\Bigg[
B_{21}h_{tx,w}''
+B_{22}h_{tx,w}' \\
&
-\left(\frac{c_1c_2\gamma}{2c_3}-B_{23}\right)a_w^{\prime}
-\left(\frac{c_2\gamma A^{\prime}}{2c_1c_3}-B_{24}\right)h_{tx,w}\Bigg]a_{-w} \,,
\end{split}
\end{equation}
where the coefficients $B_i\equiv B_i(\beta)$ contain the higher-derivative contributions.
From the variation of the action \eqref{conductivityaction1}, one can derive
the explicit form of the equations of motion for the fluctuations:
\begin{align}
\label{condeomsab}
&a_w''+\biggl(\, \sum_i \del_i\ln\gamma\cdot \phi_i'
+\left(\ln\frac{c_1 c_2}{c_3}\right)'\,\biggr)\, a_w'+\frac{c_3^2 w^2}{c_1^2}\, a_w
+ \frac{c_2^2 A'}{c_1^2}\, \left(\frac{h_{tx,w}}{c_2^2}\right)'
+ \beta  \cdot  \Big[ \ldots  \Big] = 0\,, \nonumber \\
&\left(\frac{h_{tx,w}}{c_2^2}\right)'+\frac{\gamma A'}{c_2^2}\, a_w+ \beta
\cdot  \Big[ \ldots  \Big]=0\,.
\end{align}
These equations can then be used to eliminate derivatives of the fluctuations $h_{tx,w}$ and $a_w$ that appear in the current in \eqref{currentcondI}.

To proceed, it is convenient to redefine the fluctuations in the following way, 
\begin{align} 
    h_{tx,w}&=c_2^2\left(\frac{c_1}{c_2}\right)^{-i\mathfrak{w}}\left(H_{tx,0}+i\mathfrak{w}\,H_{tx,1}\right)\,,
    \qquad a_{w}=\left(\frac{c_1}{c_2}\right)^{-i\mathfrak{w}}\left(a_{0}+i\mathfrak{w}\,a_{1}\right).\label{Capitala}
\end{align}
We impose the same boundary normalization and horizon regularity conditions (including the incoming-wave condition) as in the shear viscosity case in \eqref{boundnorm}, with the obvious replacement $H_{12,w}\longleftrightarrow a_w$.

Writing the imaginary part of the current as 
\begin{equation}
    J_1\equiv F + \beta\cdot\delta F \,,
\end{equation}
and  
recalling \eqref{kubosigma}, we can write the conductivity $\hat\sigma$,
see \eqref{defhs}, as
\begin{equation}
    \hat\sigma
    = \frac{e^2}{2\pi T}\cdot\left(F+\beta\,\delta F \right)\,.
\end{equation}
The leading contribution to the imaginary part takes the form
\begin{equation}\label{condcurrentlead}
    F=
    \frac{c_1c_2}{2c_3}\gamma\left(a_{0}a_{1}^{\prime}
    -a_{1}a_{0}^{\prime}\right)
    +\frac{1}{2c_3}a_{0}^2\gamma\left(c_1c_2^{\prime}-c_2c_1^{\prime}\right) \,,
\end{equation}
while the contribution from the  higher-derivative corrections is given by
\begin{align}
    \delta F
    &=\frac{a_0^2\left(c_2c_1^{\prime}-c_1c_2^{\prime}\right)+c_1c_2\left(a_{1}a_{0}^{\prime}-a_0a_{1}^{\prime}\right)}{6c_1^2c_2^2c_3^3}\Big(12c_1c_2c_1^{\prime}c_2^{\prime}\left(6\alpha_4+3\alpha_5+2\alpha_6+\alpha_7\right) \nonumber \\
    &+2c_1^2\big(c_2^2c_3^2(-12+V)(16\alpha_4+5\alpha_5+4\alpha_6+2\alpha_7)+6c_2^{\prime 2}(6\alpha_4+3\alpha_5+4\alpha_6+2\alpha_7)\big)\nonumber \\
    &-c_2^2A^{\prime 2}(3\gamma^2(\alpha_2+4\alpha_3)+\gamma (-4\alpha_4+4\alpha_5+20\alpha_6+10\alpha_7)+24(2\alpha_8+\alpha_9))\Big).
\end{align}
Combining these results yields the expression for the conductivity,  
\begin{equation}\label{sigmamod1ab}
\begin{split}
    \hat\sigma=\frac{e^2\gamma}{2r_h}a_0^2\bigg(&1+\beta\Bigl(\mathcal{C}\gamma(\alpha_2+4\alpha_3)+\frac{\mathcal{C}}{3}(2\alpha_4+7\alpha_5+22\alpha_6+11\alpha_7) \\
    &+\frac{\mathcal{C}}{\gamma}(16\alpha_8+8\alpha_9)-\frac{V-12}{3\gamma}(20\alpha_4+4\alpha_5+4\alpha_6+2\alpha_7)\Bigr)\bigg)\,,
\end{split}
\end{equation}
where $\calc$ is given by \eqref{defco}. 
The explicit expression for $a_0$ should come from solving the equation of motion for a specific background which often needs numerical analysis or various approximations.

To arrive at the final formula for the conductivity reported in \eqref{sigmasmod1} we eliminate $r_h$ in \eqref{sigmamod1ab}
using \eqref{Bekenstein}, \eqref{swsb},  \eqref{BGfunctions1} and  \eqref{sWsBRatio},
\begin{equation}
\frac {1}{r_h} = (\hat s)^{1/3} \left(1-\frac 13\beta\, \kappa\right)\,,\qquad {\rm where}\qquad \hat s \equiv 4G_N\, s 
\label{killrh}
\end{equation}
is the entropy density {\it per degree of freedom} of the boundary theory, \ie $\hat s\sim \frac{s}{N^2}$.

\subsubsection*{Model II}
As before, we start by evaluating the action for the model \eqref{action2} on the background with the relevant fluctuations \eqref{fluct2}. Similar to other sections, after complexifying the action, we find that the it can be rewritten as in \eqref{ConComplexL}, from which we read off the current $J_w$ and find
\begin{align}
        J_w = \, &- \frac{c_2 \gamma A'}{2 c_1 c_3}\, h_{tx,w} a_{-w} - \frac{c_1 c_2 \gamma}{2 c_3} a_w' a_{-w} - \bigg[\frac{c_2}{c_1 c_3}-\frac{4 c_2'^2}{c_1 c_2 c_3^3} \, \lambda_1 \bigg] \, h_{tx,w} h_{tx,-w}' \nonumber\\
        &- \bigg[ \frac{c_2}{2 c_1 c_3} - \frac{2 c_2'^2}{c_1 c_2 c_3^3} \, \lambda_1 + \frac{4 c_2'}{c_1 c_3^3} \, \sum_i \partial_i \lambda_1 \cdot \phi_i'  \bigg] \, h_{tx,w}'h_{tx,-w} + \frac{4 c_2'}{c_1 c_3^3} \, \lambda_1 \, h_{tx,w}' h_{tx,-w}' \nonumber\\
        &+ \bigg[ \frac{c_2 c_1'}{c_1^2 c_3} + \frac{c_2'}{c_1 c_3} - \Big(\frac{12 c_1' c_2'^2}{c_1^2 c_2 c_3^3} + \frac{4 c_2'^3}{c_1 c_2^2 c_3^3} \Big) \, \lambda_1 - \frac{4 c_2'^2}{c_1 c_2 c_3^3} \, \sum_i \partial_i \lambda_1 \cdot \phi_i'   \bigg] \, h_{tx,w} h_{tx,-w} \nonumber \\
        &+  \epsilon  \cdot  \Big[ \ldots  \Big]\,.  \label{GBConJw}
\end{align}
Using the equations of motion for the fluctuations, together with the background equations \eqref{GBBGEOMS1}, \eqref{GBBGEOMS2} and \eqref{BGEOMModelTwo}, we can show that the imaginary part of the current is indeed radially conserved, and hence can be used for calculating the retarded correlation function at the horizon.

Moreover, to further evaluate the current \eqref{GBConJw}, we use the equation of motion for $h_{tx}(r)$, and eliminate its first and higher-derivatives, \textit{i.e.} $h_{tx}'$, $h_{tx}''$, \textit{etc}. 
After using the background EoMs  \eqref{GBBGEOMS1}, \eqref{GBBGEOMS2} and \eqref{BGEOMModelTwo}, we work out the expression for the current in the hydrodynamic limit;  we find its imaginary part to be
\begin{equation}
\begin{split}
    J_1 = \frac{\gamma a_0^2}{2 c_3} \, (c_1 c_2' - c_1' c_2) + \frac{c_1 c_2 \gamma}{2 c_3} \, (a_0 a_1' - a_0' a_1) +  \epsilon  \cdot
    \Big[ \ldots  \Big]\,.
\end{split}
\end{equation}
Interestingly, we find that there is no correction to this current from the Gauss-Bonnet term, and corrections are now only from the perturbative sector, characterized by $\epsilon$.
We now use $J_1$ to calculate the conductivity $\sigma$ as in the previous section (see Model I). Since the imaginary part of the current is radially conserved, we evaluate it at the horizon, and find  to first order in $\epsilon$,
\begin{equation}
    \hat\sigma = \frac{e^2 \gamma}{2 r_h} \, a_0^2 \, \biggl(1
    -\epsilon\, \frac{4 \, (2 V-24 + \gamma \mathcal{C})^2}{3\gamma}\lambda_2 \bigg) \, ,
\end{equation}
where, as before, $\mathcal{C}$ is given by \eqref{defco}.
The final expression \eqref{sigmasmod2} is obtained replacing
$r_h$ in favor of $\hat s$, see \eqref{killrh} ---
in this model there is no difference between the
Bekenstein and the Wald entropies \eqref{sWsBModelTwo},
and thus the corresponding value of $\kappa=0$.

\section*{Acknowledgments}
We would like to thank Pavel Kovtun for valuable discussions.  AB, SC and MM
gratefully acknowledge support from the Simons Center
for Geometry and Physics, Stony Brook University, at which some of the
research for this paper was performed.  
MM acknowledges
the support of the Dr. Hyo Sang Lee Graduate Fellowship from the College of Arts and
Sciences at Lehigh University.
AB's work was supported by
NSERC through the Discovery Grants program. 
 The work of SC is supported in part by the NSF grant PHY-2210271.  
GT is supported in part by NSF grant PHY-2210271 and by the Lehigh University CORE grant with grant ID: COREAWD40.

\newpage
\appendix

\section{Appendix A}\label{app:AppendixA}
\subsection{Details of \eqref{effActionS1}}
\begin{align} \label{effBGAction1}
        \mathcal{I}^{\text{(I)}} = \, &12 c_1 c_2^3 c_3 (12 - V) - \frac{c_1 c_2^3}{2 c_3} \, \sum_i (\phi_i')^2 + \frac{c_2^3 \gamma A'^2}{2 c_1 c_3} - \frac{6 c_2^2 c_1' c_2'}{c_3} - \frac{6 c_1 c_2 c_2'^2}{c_3} + \frac{2 c_2^3 c_1' c_3'}{c_3^2} \nonumber \\
        &+ \frac{6 c_1 c_2^2 c_2' c_3'}{c_3^2} 
        - \frac{2 c_2^3 c_1''}{c_3} - \frac{6 c_1 c_2^2 c_2''}{c_3} \, ,
\end{align}
 \begin{align} \label{effBGAction2}
        \delta\mathcal{I}^{\text{(I)}} = \, &\alpha_1 \bigg( \frac{36 c_2 c_1'^2 c_2'^2}{c_1 c_3^3} + \frac{72 c_1' c_2'^3}{c_3^3} + \frac{36 c_1 c_2'^4}{c_2 c_3^3} - \frac{24 c_2^2 c_1'^2 c_2' c_3'}{c_1 c_3^4} - \frac{96 c_2 c_1' c_2'^2 c_3'}{c_3^4} - \frac{72 c_1 c_2'^3 c_3'}{c_3^4}  \nonumber \\ 
        &+ \frac{4 c_2^3 c_1'^2 c_3'^2}{c_1 c_3^5} + \frac{24 c_2^2 c_1' c_2' c_3'^2}{c_3^5} + \frac{36 c_1 c_2 c_2'^2 c_3'^2}{c_3^5} + \frac{24 c_2^2 c_1' c_2' c_1''}{c_1 c_3^3} + \frac{24 c_2 c_2'^2 c_1''}{c_3^3} - \frac{8 c_2^3 c_1' c_3' c_1''}{c_1 c_3^4} \nonumber \\
        & - \frac{24 c_2^2 c_2' c_3' c_1''}{c_3^4} + \frac{4 c_2^3 c_1''^2}{c_1 c_3^3} + \frac{72 c_2 c_1' c_2' c_2''}{c_3^3} + \frac{72 c_1 c_2'^2 c_2''}{c_3^3} - \frac{24 c_2^2 c_1' c_3' c_2''}{c_3^4} -\frac{72 c_1 c_2 c_2' c_3' c_2''}{c_3^4} \nonumber \\ 
        &+ \frac{24 c_2^2 c_1'' c_2''}{c_3^3} + \frac{36 c_1 c_2 c_2''^2}{c_3^3} \bigg) +  \alpha_2 \bigg( \frac{12 c_2 c_1'^2 c_2'^2}{c_1 c_3^3} + \frac{12 c_1' c_2'^3}{c_3^3} +\frac{12 c_1 c_2'^4}{c_2 c_3^3} - \frac{6 c_2^2 c_1'^2 c_2' c_3'}{c_1 c_3^4} \nonumber \\
        &- \frac{6 c_2 c_1' c_2'^2 c_3'}{c_3^4} - \frac{12 c_1 c_2'^3 c_3'}{c_3^4} + \frac{2 c_2^3 c_1'^2 c_3'^2}{c_1 c_3^5} + \frac{6 c_2^2 c_1' c_2' c_3'^2}{c_3^5} + \frac{12 c_1 c_2 c_2'^2 c_3'^2}{c_3^5} + \frac{6 c_2^2 c_1' c_2' c_1''}{c_1 c_3^3} \nonumber \\
        &- \frac{4 c_2^3 c_1' c_3' c_1''}{c_1 c_3^4} - \frac{6 c_2^2 c_2' c_3' c_1''}{c_3^4} + \frac{2 c_2^3 c_1''^2}{c_1 c_3^3} + \frac{6 c_2 c_1' c_2' c_2''}{c_3^3} + \frac{12 c_1 c_2'^2 c_2''}{c_3^3} - \frac{6 c_2^2 c_1' c_3' c_2''}{c_3^4} \nonumber \\
        & -\frac{24 c_1 c_2 c_2' c_3' c_2''}{c_3^4} + \frac{6 c_2^2 c_1'' c_2''}{c_3^3} + \frac{12 c_1 c_2 c_2''^2}{c_3^3} \bigg) + \alpha_3 \bigg( \frac{12 c_2 c_1'^2 c_2'^2}{c_1 c_3^3} + \frac{12 c_1 c_2'^4}{c_2 c_3^3} + \frac{4 c_2^3 c_1'^2 c_3'^2}{c_1 c_3^5} \nonumber \\
        &+ \frac{12 c_1 c_2 c_2'^2 c_3'^2}{c_3^5} - \frac{8 c_2^3 c_1' c_3' c_1''}{c_1 c_3^4} + \frac{4 c_2^3 c_1''^2}{c_1 c_3^3} - \frac{24 c_1 c_2 c_2' c_3' c_2''}{c_3^4} + \frac{12 c_1 c_2 c_2''^2}{c_3^3} \bigg) \nonumber \\ 
        &+\alpha_4 \bigg( \frac{12 c_2^2 c_1' c_2' A'^2}{c_1^2 c_3^3} + \frac{12 c_2 c_2'^2 A'^2}{c_1 c_3^3} - \frac{4 c_2^3 c_1' c_3' A'^2}{c_1^2 c_3^4} - \frac{12 c_2^2 c_2' c_3' A'^2}{c_1 c_3^4} + \frac{4 c_2^3 c_1'' A'^2}{c_1^2 c_3^3} \nonumber \\
        &+ \frac{12 c_2^2 c_2'' A'^2}{c_1 c_3^3}\bigg) + \alpha_5 \bigg(\frac{3 c_2^2 c_1' c_2' A'^2}{c_1^2 c_3^3} - \frac{2 c_2^3 c_1' c_3' A'^2}{c_1^2 c_3^4} - \frac{3 c_2^2 c_2' c_3' A'^2}{c_1 c_3^4}  + \frac{2 c_2^3 c_1'' A'^2}{c_1^2 c_3^3}\nonumber \\
        &+ \frac{3 c_2^2 c_2'' A'^2}{c_1 c_3^3}\bigg) + (2 \alpha_6 + \alpha_7) \bigg(\frac{2 c_2^3 c_1'' A'^2}{c_1^2 c_3^3} - \frac{2 c_2^3 c_1' c_3' A'^2}{c_1^2 c_3^4} \bigg) + (2 \alpha_8 + \alpha_9) \frac{2 c_2^3 A'^4}{c_1^3 c_3^3} \, .
\end{align}
\newpage
\subsection{Details of \eqref{BGTotalDer}}
 \begin{align} \label{DeltaBMod1}
        \delta_{\mathcal{B}}^{\text{(I)}} = \, &\alpha_1 \bigg( \frac{48 c_2^2 c_1'^2 c_2'}{c_1 c_3^3} + \frac{48 c_2 c_1' c_2'^2}{c_3^3} + \frac{48 c_1 c_2'^3}{c_3^3} - \frac{16 c_2^3 c_1'^2 c_3'}{c_1 c_3^4} + \frac{24 c_2^2 c_1' c_2' c_3'}{c_3^4} + \frac{24 c_1 c_2 c_2'^2 c_3'}{c_3^4}\nonumber \\
        & - \frac{24 c_2^3 c_1' c_3'^2}{c_3^5} -\frac{72 c_1 c_2^2 c_2' c_3'^2}{c_3^5} + \frac{16 c_2^3 c_1' c_1''}{c_1 c_3^3} - \frac{24 c_2^2 c_2' c_1''}{c_3^3} + \frac{24 c_2^3 c_3' c_1''}{c_3^4} - \frac{24 c_1 c_2 c_2' c_2''}{c_3^3}\nonumber \\ 
        &+ \frac{72 c_1 c_2^2 c_3' c_2''}{c_3^4} + \frac{8 c_2^3 c_1' c_3''}{c_3^4} + \frac{24 c_1 c_2^2 c_2' c_3''}{c_3^4} - \frac{8 c_2^3 c_1'''}{c_3^3} - \frac{24 c_1 c_2^2 c_2'''}{c_3^3}\bigg) + \alpha_2 \bigg( \frac{12 c_2^2 c_1'^2 c_2'}{c_1 c_3^3}\nonumber \\ 
        &+ \frac{12 c_2 c_1' c_2'^2}{c_3^3} + \frac{12 c_1 c_2'^3}{c_3^3} - \frac{8 c_2^3 c_1'^2 c_3'}{c_1 c_3^4}+ \frac{12 c_2^2 c_1' c_2' c_3'}{c_3^4} + \frac{6 c_1 c_2 c_2'^2 c_3'}{c_3^4} - \frac{12 c_2^3 c_1' c_3'^2}{c_3^5}\nonumber \\ 
        &- \frac{18 c_1 c_2^2 c_2' c_3'^2}{c_3^5} + \frac{8 c_2^3 c_1' c_1''}{c_1 c_3^3} - \frac{12 c_2^2 c_2' c_1''}{c_3^3} + \frac{12 c_2^3 c_3' c_1''}{c_3^4} - \frac{6 c_1 c_2 c_2' c_2''}{c_3^3} + \frac{18 c_1 c_2^2 c_3' c_2''}{c_3^4} \nonumber \\
        &+ \frac{4 c_2^3 c_1' c_3''}{c_3^4} + \frac{6 c_1 c_2^2 c_2' c_3''}{c_3^4} - \frac{4 c_2^3 c_1'''}{c_3^3} - \frac{6 c_1 c_2^2 c_2'''}{c_3^3} \bigg) + \alpha_3 \bigg( \frac{24 c_2 c_1' c_2'^2}{c_3^3} - \frac{16 c_2^3 c_1'^2 c_3'}{c_1 c_3^4}\nonumber \\ 
        &+ \frac{24 c_2^2 c_1' c_2' c_3'}{c_3^4} - \frac{24 c_2^3 c_1' c_3'^2}{c_3^5} + \frac{16 c_2^3 c_1' c_1''}{c_1 c_3^3} - \frac{24 c_2^2 c_2' c_1''}{c_3^3}+ \frac{24 c_2^3 c_3' c_1''}{c_3^4} + \frac{8 c_2^3 c_1' c_3''}{c_3^4} - \frac{8 c_2^3 c_1'''}{c_3^3}\bigg) \nonumber \\  
        &+\alpha_4 \bigg( \frac{12 c_2^3 c_1' A'^2}{c_1^2 c_3^3} + \frac{8 c_2^3 c_3' A'^2}{c_1 c_3^4} - \frac{8 c_2^3 A' A''}{c_1 c_3^3} \bigg) + \alpha_5 \bigg( \frac{6 c_2^3 c_1' A'^2}{c_1^2 c_3^3} - \frac{3 c_2^2 c_2' A'^2}{c_1 c_3^3} \nonumber \\
        &+ \frac{4 c_2^3 c_3' A'^2}{c_1 c_3^4} - \frac{4 c_2^3 A' A''}{c_1 c_3^3} \bigg) + (2 \alpha_6 + \alpha_7) \bigg( \frac{6 c_2^3 c_1' A'^2}{c_1^2 c_3^3} - \frac{6 c_2^2 c_2' A'^2}{c_1 c_3^3} + \frac{4 c_2^3 c_3' A'^2}{c_1 c_3^4}\nonumber \\
        &- \frac{4 c_2^3 A' A''}{c_1 c_3^3} \bigg) \, .
\end{align}
\vspace{0.1cm}
\subsection{EoMs for the background fields in Model II}
\begin{align} \label{BGEOMModelTwo}
        &- \frac{6 c_2^2 c_2''}{c_3} + c_2^3 c_3 (12 - V) - \frac{c_2^3}{2 c_3} \sum_i (\phi_i')^2 - \frac{c_2^3 \gamma A'^2}{2 c_1^2 c_3} - \frac{6 c_2 c_2'^2}{c_3} + \frac{6 c_2^2 c_2' c_3'}{c_3^2} \nonumber \\
        &+\lambda_1 \, \frac{24 c_2'^2 }{c_3^4}\, (c_2'' c_3 - c_2' c_3' )  +\frac{24 c_2'}{c_3^4} \, \sum_i (\partial_i \lambda_1) \cdot   (c_3 c_2'^2 \phi_i + 2 c_3 c_2 c_2'' \phi_i + c_2 c_3 c_2' \phi_i'' - 3 c_2 c_2' c_3' \phi_i) \nonumber \\
        & +  \frac{24 c_2 c_2'^2}{c_3^3} \, \sum_{i,j} (\partial_{ij}^2 \lambda_1) \cdot \phi_i' \phi_j' + \epsilon  \cdot  \Big[ \ldots  \Big] =0 \, ,   \nonumber \displaybreak \\
        &- \frac{6 c_2^2 c_1''}{c_3} - \frac{12 c_1 c_2 c_2''}{c_3} + 3 c_1 c_2^2 c_3 (12 - V) - \frac{3 c_1 c_2^2}{2 c_3} \sum_i (\phi_i')^2 + \frac{3 c_2^2 \gamma A'^2}{2 c_1 c_3} - \frac{12 c_2 c_1' c_2'}{c_3} - \frac{6 c_1 c_2'^2}{c_3} \nonumber \\
        &+ \frac{6 c_2^2 c_1' c_3'}{c_3^2} + \frac{12 c_1 c_2 c_2' c_3'}{c_3^2} + \lambda_1 \, \frac{24 c_2'}{c_3^4} \, (c_3 c_2' c_1'' - 3 c_2' c_1' c_3' + 2 c_3 c_1' c_2'') \nonumber \\
        &+ \frac{24}{c_3^4} \, \sum_i (\partial_i \lambda_1) \cdot (3 c_3 c_1' c_2'^2 \phi_i - 6 c_2 c_1' c_2' c_3' \phi_i' - 3 c_1 c_2'^2 c_3' \phi_i' + 2 c_2 c_3 c_2' c_1'' \phi_i' + 2 c_2 c_3 c_1' c_2'' \phi_i'  \nonumber \\
        &+ 2 c_1 c_3 c_2' c_2'' \phi_i' + 2 c_2 c_3 c_1' c_2' \phi_i'' + c_1 c_3 c_2'^2 \phi_i'') + \frac{24 c_2'}{c_3^3} \, (2 c_2 c_1' + c_1 c_2') \, \sum_{i,j} (\partial_{ij}^2 \lambda_1) \cdot \phi_i' \phi_j'  \nonumber \\
        &+ \epsilon  \cdot  \Big[ \ldots  \Big] =0 \, , \nonumber \\
        &c_1 c_2^3 (12 - V) + \frac{c_1 c_2^3}{2 c_3^2} \sum_i (\phi_i')^2 - \frac{c_2^3 \gamma A'^2}{2 c_1 c_3^2} - \frac{6 c_2^2 c_1' c_2'}{c_3^2} - \frac{6 c_1 c_2 c_2'^2}{c_3^2} + \lambda_1 \, \frac{24 c_1' c_2'^3}{c_3^4} \nonumber \\
        &+\frac{24 c_2'^2 (3 c_2 c_1' + c_1 c_2')}{c_3^4} \, \sum_i (\partial_i \lambda_1) \cdot \phi_i'  + \epsilon  \cdot  \Big[ \ldots  \Big] =0 \, , \nonumber \\ 
        &\frac{c_1 c_2^3 \phi_i''}{c_3} - c_1 c_2^3 c_3 (\partial_i V) + \frac{c_2^3 A'^2 (\partial_i \gamma)}{2 c_1 c_3} + \frac{c_2^3 c_1' \phi_i'}{c_3} + \frac{3 c_1 c_2^2 c_2' \phi_i'}{c_3} - \frac{c_1 c_2^3 c_3' \phi_i'}{c_3^2} + \frac{24 c_2'}{c_3^4} (c_3 c_1' c_2'^2 \nonumber \\
        &- 3 c_2 c_1' c_2' c_3' - c_1 c_2'^2 c_3' + c_2 c_3 c_2' c_1'' + 2 c_2 c_3 c_1' c_2'' + c_1 c_3 c_2' c_2'') \, (\partial_i \lambda_1)   + \epsilon  \cdot  \Big[ \ldots  \Big] =0 \, .
\end{align}
\vspace{0.001cm}
\subsection{Coefficients of the current in \texorpdfstring{\eqref{shearcurrent}}{(shearcurrent)}}\label{app:AppendixAI}
The coefficients $B_i$ should all be multiplied by $\beta$:
\begin{align}
    A_1=&\frac{c_1}{c_2c_3}\,,
    A_2=-\frac{2c_1c_2'}{c_2^2c_3}\,, \quad B_2=\frac{a_2}{2c_2c_3^3}\left(c_1^{\prime}
    -\frac{c_1c_2'}{c_2}
    -\frac{c_1c_3'}{c_3}\right)
    -\frac{\alpha_3 c_1}{c_2c_3^3}\left(\frac{4c_2'}{c_2}
    +\frac{2c_3'}{c_3}\right)\, , \nonumber \\
    B_1=&\frac{\alpha_2}{2c_2c_3^3}+\frac{2c_1\alpha_3}{c_2c_3^3}\,,\quad B_3=\frac{\alpha_2}{2c_1c_2c_3}\,,\quad B_5=-\frac{4\alpha_3}{c_1c_2c_3}-\frac{\alpha_2}{2c_1c_2c_3}\,, \nonumber \\
    B_4=&
    \alpha_1\Bigg(-\frac{12c_1'c_2'}{c_2^2c_3^3}-\frac{12c_1c_2'^2}{c_2^3c_3^3}+\frac{4c_1'c_3'}{c_2c_3^4}+\frac{12c_1c_2'c_3'}{c_2^2c_3^4}-\frac{4c_1''}{c_2c_3^3}-\frac{12c_1c_2''}{c_2^2c_3^3}\Bigg) \nonumber \\
    &+\alpha_2\Bigg(-\frac{2c_1'c_2'}{c_2^2c_3^3}-\frac{2c_1c_2'^2}{c_2^3c_3^3}+\frac{c_1'c_3'}{c_2c_3^4}+\frac{5c_1c_2'c_3'}{c_2^2c_3^4}-\frac{c_1''}{c_2c_3^3}-\frac{5c_1c_2''}{c_2^2c_3^3}\Bigg) \nonumber \\
    &+\alpha_3\Bigg(\frac{4c_1c_2'^2}{c_2^3c_3^3}+\frac{8c_1c_2'c_3'}{c_2^2c_3^4}-\frac{8c_1c_2''}{c_2^2c_3^3}\Bigg)
    -\frac{2\alpha_4A'^2}{c_1c_2c_3^3}-\frac{\alpha_5A'^2}{2c_1c_2c_3^3} \, , \nonumber\\
    B_6=&-\frac{\alpha_4A'^2}{c_1c_2c_3^3}
    +\phi'\Bigg(-\frac{c_1'\alpha_2'}{2c_2c_3^3}+\frac{c_1c_2'\alpha_2'}{2c_2^2c_3^3}+\frac{c_1c_3'\alpha_2'}{2c_2c_3^4}+\frac{4c_1c_2'\alpha_3'}{c_2^2c_3^3}+\frac{2c_1c_3'\alpha_3'}{c_2c_3^4}\Bigg)\nonumber\\
    &+\alpha_1\Bigg(-\frac{6c_1'c_2'}{c_2^2c_3^3}-\frac{6c_1c_2'^2}{c_2^3c_3^3}+\frac{2c_1'c_3'}{c_2c_3^4}+\frac{6c_1c_2'c_3'}{c_2^2c_3^4}-\frac{2c_1''}{c_2c_3^3}-\frac{6c_1c_2''}{c_2^2c_3^3}\Bigg) \nonumber
    \\
    &+\alpha_2\Bigg(\frac{c_1'^2}{2c_1c_2c_3^3}-\frac{5c_1c_2'^2}{2c_2^3c_3^3}+\frac{c_1'c_3'}{c_2c_3^4}-\frac{c_1c_2'c_3'}{c_2^2c_3^4}-\frac{3c_1c_3'^2}{2c_2c_3^5}-\frac{c_1''}{2c_2c_3^3}+\frac{c_1c_2''}{2c_2^2c_3^3}+\frac{c_1c_3''}{2c_2c_3^4}\Bigg)\nonumber\\
    &+\alpha_3\Bigg(\frac{2c_1'^2}{c_1c_2c_3^3}+\frac{4c_1'c_2'}{c_2^2c_3^3}-\frac{4c_1c_2'^2}{c_2^3c_3^3}+\frac{2c_1'c_3'}{c_2c_3^4}-\frac{10c_1c_2'c_3'}{c_2^2c_3^4}-\frac{6c_1c_3'^2}{c_2c_3^5}+\frac{8c_1c_2''}{c_2^2c_3^3}+\frac{2c_1c_3''}{c_2c_3^4}\Bigg)\, ,\nonumber\\
    B_7=&
    -\alpha_2'\frac{\phi'}{2c_1c_2c_3}
    +\alpha_2\frac{c_1'}{c_1^2c_2c_3}
    +\alpha_3\left(\frac{6c_1'}{c_1^2c_2c_3}
    +\frac{4c_2'}{c_1c_2^2c_3}\right) \, ,  \nonumber\\
    B_8=&\phi'\Bigg(\frac{12c_1'c_2'\alpha_1'}{c_2^2c_3^3}+\frac{12c_1c_2'^2\alpha_1'}{c_2^3c_3^3}-\frac{4c_1'c_3'\alpha_1'}{c_2c_3^4}-\frac{12c_1c_2'c_3'\alpha_1'}{c_2^2c_3^4}+\frac{2c_1'c_2'\alpha_2'}{c_2^2c_3^3}+\frac{2c_1c_2'^2\alpha_2'}{c_2^3c_3^3}\nonumber\\
    &\quad\quad\quad-\frac{c_1'c_3'\alpha_2'}{c_2c_3^4}-\frac{5c_1c_2'c_3'\alpha_2'}{c_2^2c_3^4}-\frac{4c_1c_2'^2\alpha_3'}{c_2^3c_3^3}-\frac{8c_1c_2'c_3'\alpha_3'}{c_2^2c_3^4}+\frac{2\alpha_4'A'^2}{c_1c_2c_3^3}+\frac{\alpha_5'A'^2}{2c_1c_2c_3^3}\nonumber\\
    &\quad\quad\quad
    +\frac{4\alpha_1'c_1''}{c_2c_3^3}
    +\frac{\alpha_2'c_1''}{c_2c_3^3}
    +\frac{12\alpha_1'c_1c_2''}{c_2^2c_3^3}
    +\frac{5c_1\alpha_2'c_2''}{c_2^2c_3^3}
    +\frac{8c_1\alpha_3'c_2''}{c_2^2c_3^3}
    \Bigg)\nonumber\\
    &
    +\alpha_1\Bigg(-\frac{12c_1'^2c_2'}{c_1c_2^2c_3^3}+\frac{12c_1'c_2'^2}{c_2^3c_3^3}+\frac{4c_1'^2c_3'}{c_1c_2c_3^4}-\frac{32c_1'c_2'c_3'}{c_2^2c_3^4}-\frac{36c_1c_2'^2c_3'}{c_2^3c_3^4}+\frac{12c_1'c_3'^2}{c_2c_3^5}\nonumber\\
    &\quad\quad\quad
    +\frac{36c_1c_2'c_3'^2}{c_2^2c_3^5}-\frac{4c_1'c_1''}{c_1c_2c_3^3}+\frac{20c_2'c_1''}{c_2^2c_3^3}-\frac{12c_3'c_1''}{c_2c_3^4}+\frac{12c_1'c_2''}{c_2^2c_3^3}+\frac{36c_1c_2'c_2''}{c_2^3c_3^3}
    \nonumber\\
    &\quad\quad\quad
    -\frac{36c_1c_3'c_2''}{c_2^2c_3^4}-\frac{4c_1'c_3''}{c_2c_3^4}-\frac{12c_1c_2'c_3''}{c_2^2c_3^4}+\frac{4c_1^{\prime\prime\prime}}{c_2c_3^3}+\frac{12c_1c_2^{\prime\prime\prime}}{c_2^2c_3^3}\Bigg) 
    \nonumber\\
    &
    +\alpha_2\Bigg(-\frac{5c_1'^2c_2'}{c_1c_2^2c_3^3}-\frac{c_1'c_2'^2}{c_2^3c_3^3}-\frac{2c_1c_2'^3}{c_2^4c_3^3}+\frac{c_1'^2c_3'}{c_1c_2c_3^4}-\frac{10c_1'c_2'c_3'}{c_2^2c_3^4}-\frac{9c_1c_2'^2c_3'}{c_2^3c_3^4}+\frac{3c_1'c_3'^2}{c_2c_3^5}
    \nonumber\\
    &\quad\quad\quad
    +\frac{15c_1c_2'c_3'^2}{c_2^2c_3^5}-\frac{c_1'c_1''}{c_1c_2c_3^3}+\frac{5c_2'c_1''}{c_2^2c_3^3}-\frac{3c_3'c_1''}{c_2c_3^4}+\frac{5c_1'c_2''}{c_2^2c_3^3}+\frac{9c_1c_2'c_2''}{c_2^3c_3^3}-\frac{15c_1c_3'c_2''}{c_2^2c_3^4}
    \nonumber\\
    &\quad\quad\quad
    -\frac{c_1'c_3''}{c_2c_3^4}-\frac{5c_1c_2'c_3''}{c_2^2c_3^4}+\frac{c_1^{\prime\prime\prime}}{c_2c_3^3}+\frac{5c_1c_2^{\prime\prime\prime}}{c_2^2c_3^3}\Bigg)
    \nonumber\\
    &
    +\alpha_3\Bigg(-\frac{8c_1'^2c_2'}{c_1c_2^2c_3^3}-\frac{4c_1'c_2'^2}{c_2^3c_3^3}-\frac{4c_1c_2'^3}{c_2^4c_3^3}-\frac{8c_1'c_2'c_3'}{c_2^2c_3^4}+\frac{24c_1c_2'c_3'^2}{c_2^2c_3^5}+\frac{8c_1'c_2''}{c_2^2c_3^3}-\frac{24c_1c_3'c_2''}{c_2^2c_3^4}
    \nonumber\\
    &\quad\quad\quad
    -\frac{8c_1c_2'c_3''}{c_2^2c_3^4}+\frac{8c_1c_2^{\prime\prime\prime}}{c_2^2c_3^3}\Bigg)
    +\alpha_5\Bigg(-\frac{c_1'A'^2}{c_1^2c_2c_3^3}+\frac{3c_2'A'^2}{2c_1c_2^2c_3^3}-\frac{c_3'A'^2}{c_1c_2c_3^4}+\frac{A'A''}{c_1c_2c_3^3}\Bigg)
    \nonumber\\
    &
    +\alpha_4\Bigg(-\frac{4c_1'A'^2}{c_1^2c_2c_3^3}+\frac{4c_2'A'^2}{c_1c_2^2c_3^3}-\frac{4c_3'A'^2}{c_1c_2c_3^4}+\frac{4A'A''}{c_1c_2c_3^3}\Bigg)\,.
\end{align}

\section{Appendix B -- Bulk Viscosity of the Model  \cite{Buchel:2010gd} }\label{app:AppendixB}

The first computation of the bulk viscosity in non-conformal
charged plasma was performed in \cite{Buchel:2010gd},
computing the sound channel quasinormal modes of the charged
black branes, and extracting the bulk viscosity
from their attenuation. This result was reproduced from the null
horizon focusing equation in \cite{Eling:2011ms}. 
We illustrate how the framework of section \ref{TransportCoefficients}
can be applied here.

The holographic model of \cite{Buchel:2010gd} is represented
by the following effective action
\begin{align}
S_5
=&\frac{1}{16\pi G_5} \int_{\calm_5}d^5\xi \sqrt{-g}\biggr(R-\frac 14 \phi^{4/3} F^2 -
\frac{1}{3}\phi^{-2}\left(\del\phi\right)^2+4\phi^{2/3}+8\phi^{-1/3}
-\frac 12 \left(\del\chi\right)^2 \nonumber \\
&-\frac{m^2}{2}\chi^2+\calo\left(\chi^4\right)
\biggr)\,,
\label{ac5ab}
\end{align}
where the bulk scalar $\chi$ is treated to quadratic order,
correspondingly, representing a single $U(1)$ charged
STU conformal model \cite{Behrndt:1998jd}, with softly broken
scale invariance by an operator of dimension
\begin{equation}
\Delta (\Delta-4)=m^2\,.
\label{md}
\end{equation} 
In  \cite{Buchel:2010gd} the choice was made for a mass deformation with $\Delta=3$ ($m^2=-3$).

From the perspective of the effective action \eqref{2der},
we have two bulk scalars $\phi_1,\phi_2$,
\begin{equation}
\phi_1\equiv \sqrt{\frac 23}\ln\phi\,,\qquad \phi_2\equiv \chi\,,
\label{scalarsab}
\end{equation}
with
\begin{equation}
\gamma\{\phi_i\}\equiv \exp\left\{{\sqrt \frac{8}{3}\phi_1}\right\}\,,
\quad V\{\phi_i\}\equiv 12-4 \exp\left\{{\sqrt \frac{2}{3}\phi_1}\right\}
-8 \exp\left\{{-\sqrt \frac{1}{6}\phi_1}\right\}-\frac 32 \phi_2^2\,.
\label{gammapot}
\end{equation}
The scalar $\phi_1$ does not have any non-normalizable mode turned on,
so $\lambda_1=0$ in \eqref{zibads}, and thus $z_{1,0}\equiv 0$, and the
bulk viscosity comes entirely from the scalar $z_{2,0}$, see
\eqref{zetasmod2},
\begin{equation}
\frac{\zeta}{s}=\frac{1}{9\pi}\, z_{2,0}^2\bigg|^{horizon}\qquad
\Longrightarrow\qquad  \frac{\zeta}{\eta}=\frac{4}{9}\, z_{2,0}^2\bigg|^{horizon}\,,
\label{zetaab}
\end{equation}
where in the second equality we have used the universal shear viscosity
of the charged plasma \cite{Benincasa:2006fu}.

We are interested in the bulk viscosity from the softly broken scale invariance
of the STU model. Thus, we can treat the scalar $\phi_2$, and its
gauge invariant fluctuation $z_{2,0}$, in the probe approximation.
In the probe approximation the black brane solution of the STU model
\cite{Behrndt:1998jd} is known analytically, see \eqref{BGEOMS3},
\begin{equation}
\begin{split}
&c_1=\frac{\mu\,\sqrt{(1+q)}\,\sqrt{F}}{\sqrt q\, r\, H}\,,\qquad
c_2=\frac{\mu\,\sqrt{(1+q)}\,\sqrt{H}}{\sqrt q\, r}\,,\qquad
c_3=\frac{\sqrt{H}}{r\, \sqrt{F}}\,,\\
&A=\mu\,\frac{1-r^2}{H^3}\,,\qquad \phi_1=\sqrt 6\, \ln  H\,,
\end{split}
\label{stuab}
\end{equation}
where
\begin{equation}
F\equiv (1-r^2)\,(1+(1+q)\,r^2)\,,\qquad H\equiv (1+q\, r^2)^{1/3}\,.
\label{fhab}
\end{equation}
In \eqref{stuab} $\mu$ is a $U(1)$-charge chemical potential, 
and from \eqref{Temp} we find
\begin{equation}
\frac{2\pi T}{\mu} = \frac{q+2}{\sqrt{q}}\,.
\label{tmuab}
\end{equation}
Note that the minimal (critical) temperature $T_c$ occurs when $q=2$,
\begin{equation}
\frac{2\pi T_{c}}{\mu} =2\sqrt 2\,.
\label{tmuabc}
\end{equation}
Assuming that the near-AdS-boundary expansion of $\phi_2$
takes the form (see \eqref{nonnorm})
\begin{equation}
\phi_2=\lambda_2\cdot r^{4-3}+\cdots\,,\qquad {\rm as}\  r\to 0 \,,
\label{nonnormab}
\end{equation}
we numerically solve \eqref{BulkEq2} (with $\beta=0$),
in the hydrodynamic $w=0$ limit,
and  with the near-AdS-boundary asymptotic of $z_{2,0}$, see \eqref{zibads},
as 
\begin{equation}
z_{2,0}=\lambda_2\cdot \frac{4-3}{2} \cdot r^{4-3}+\cdots\,,\qquad {\rm as}\  r\to 0 \,.
\label{nonnormabz}
\end{equation}

\begin{figure}[htb!]
\begin{center}
\includegraphics[width=10cm, angle=0]{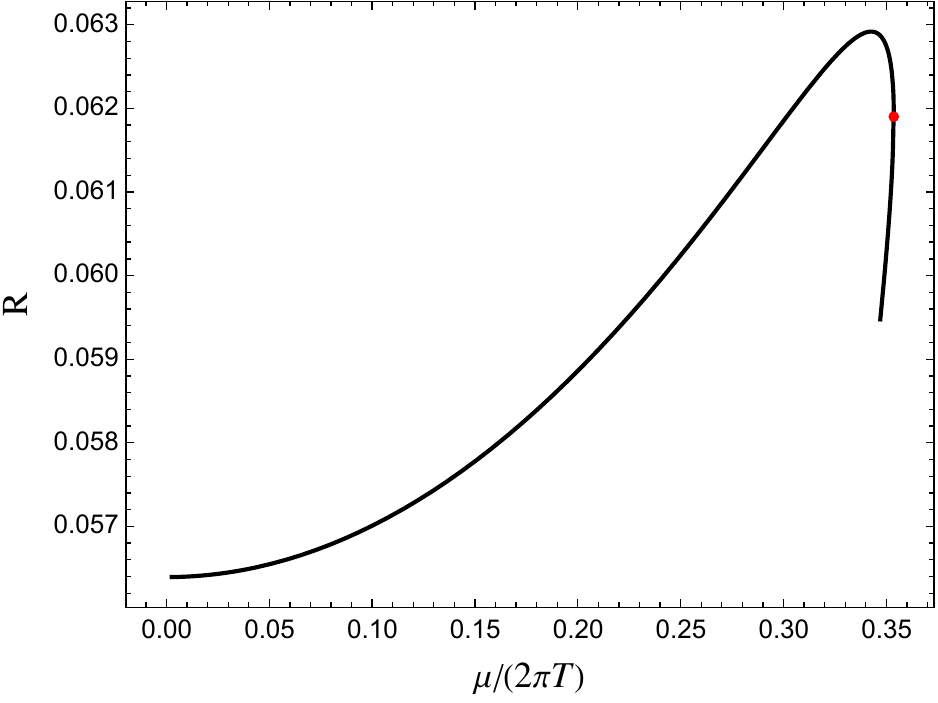}
\caption{The ratio $R$ of the bulk viscosity to the shear viscosity in the
charged hQGP of \cite{Buchel:2010gd}, see \eqref{deflab},
as a function of $\frac{\mu}{2\pi T}$. The red dot represents the
computation in \cite{Buchel:2010gd}, at the critical temperature \eqref{tmuabc}.}
\label{figure1}
\end{center}
\end{figure}

The bulk viscosity in \cite{Buchel:2010gd} (see eq.~(3.75) there)
is parametrized as 
\begin{equation}
\frac{\zeta}{\eta}=R \left(\frac T\mu\right)\,\, \lambda^2+\calo(\lambda^4)\,,
\label{deflab}
\end{equation}
where
$R$ is a function of $\frac{T}{\mu}$ and 
$\lambda$ is defined as
\begin{equation}
\lambda= \lambda_2\cdot \left(\frac {2}{1+q}\right)^{1/4}\,.
\label{deflab2}
\end{equation}
Using \eqref{zetaab}, \eqref{tmuab} and \eqref{deflab2},
we present in fig.~\ref{figure1} a plot of $R$ 
versus $\frac{\mu}{2\pi T}$. 
In \cite{Buchel:2010gd} ${R}$ was evaluated only at the 
critical temperature --- it agrees with the corresponding value on the plot
fig.~\ref{figure1}, the red dot, with an accuracy of less than $0.1\%$.

\pagebreak

\bibliographystyle{JHEP}
\bibliography{cthd}

\end{document}